\documentclass[aps,prd,reprint,groupedaddress,amsmath,amssynb,nofootinbib]{revtex4-2}
\usepackage{color, graphicx}
\usepackage[colorlinks=true,linkcolor=blue,citecolor=blue, urlcolor=blue]{hyperref}
\usepackage{fancybox, color, float, bm,braket}
\allowdisplaybreaks[1]

\begin{document}
\title{Simulation of Lindblad equations for quarkonium in the quark-gluon plasma}
\author{Takahiro Miura}
 \email{takahiro.physics@gmail.com}
\author{Yukinao Akamatsu}
 \email{yukinao.a.phys@gmail.com}
\author{Masayuki Asakawa}
 \email{yuki@phys.sci.osaka-u.ac.jp}
\author{Yukana Kaida}
\affiliation{Department of Physics, Osaka University, Toyonaka, Osaka, 560-0043, Japan}
\date{\today}

\begin{abstract}
We study the properties of the Lindbladian quantum mechanical evolution of quarkonia with non-Abelian charges (color-singlet and octet) in the quark-gluon plasma.
We confirm that heavy quark recoils in the Lindblad equation correctly thermalize quarkonium colorful states within statistical errors from the simulation method.
We also demonstrate that the Lindblad equation in the dipole limit can provide an efficient alternative method, which is applicable to a finite time evolution before thermalization and dramatically reduces the numerical cost.
Our findings will serve as a foundation for large-scale simulation of quarkonium dynamics in the relativistic heavy-ion collisions.
\end{abstract}

\maketitle

\section{Introduction}
Understanding the nature of interacting systems of quarks and gluons has been a fundamental challenge for decades.
Even though it is established that the microscopic interaction is governed by the Quantum ChromoDynamics (QCD), their physical behavior at the macroscopic scale involves non-perturbative quantum fluctuations and leads to nontrivial phenomena such as quark confinement.
At extremely high temperatures, the thermal fluctuations dominate at short distances and screen the confining force at long distances so that the quarks and gluons behave as effective degrees of freedom.
Around the transition temperature of this deconfinement, our knowledge about the matter, the Quark-Gluon Plasma (QGP) \cite{Yagi:985487}, is still limited because of the strongly coupled nature of its constituents.

The nature of the QGP has been studied by its behaviors at low energy and momentum.
At such scales, the microscopic information on how quarks and gluons interact with each other is integrated into macroscopic physical quantities such as pressure and transport coefficients.
Hydrodynamic collective behavior observed in nuclear collisions at ultra-relativistic energies can be used to constrain the shear viscosity $\eta$ of the QGP.
Its small value $\eta/s \sim 0.1-0.2$ \cite{Bernhard:2019bmu} in the unit of entropy density $s$ hinted at the realization of a strongly coupled system and early thermalization.

Experimental measurements of heavy quark bound states, in particular the clear spectral observations of bottomonium states $\Upsilon$ through dimuon pairs at the Large Hadron Collider (LHC) \cite{Chatrchyan:2011pe, Chatrchyan:2012lxa, Khachatryan:2016xxp}, opened up a new opportunity to study QGP.
Description of quarkonium as an open quantum system \cite{BRE02, rivas2012open} was motivated by the discovery of the in-medium complex potential \cite{Laine:2006ns, Beraudo:2007ky, Brambilla:2008cx, Rothkopf:2011db}.
Since then, it has been pioneered by \cite{Young:2010jq, Borghini:2011yq, Borghini:2011ms, Akamatsu:2011se, Blaizot:2015hya, Blaizot:2017ypk, Blaizot:2018oev} and further developed to the present form --- the Lindblad equations \cite{Lindblad:1975ef, Gorini:1975nb} in several regimes, which are now available \cite{Akamatsu:2014qsa, Brambilla:2016wgg, Brambilla:2017zei, Yao:2018nmy} and can be simulated \cite{Rothkopf:2013kya, Kajimoto:2017rel, Akamatsu:2018xim, Miura:2019ssi, Alund:2020ctu, Akamatsu:2021vsh,  Brambilla:2020qwo, Brambilla:2021wkt, Brambilla:2022ynh, Sharma:2019xum, Yao:2020xzw} (see also \cite{Akamatsu:2020ypb, Yao:2021lus, Sharma:2021vvu} for reviews).
In contrast to the hydrodynamics, quarkonium acts as a slow dynamical probe that locally couples to gluons.
In this respect, the Lindblad equations are characterized by the correlations of color electric fields at short distances and long time scales.
These correlation functions are so-called in-medium potential and momentum diffusion constant from the viewpoint of quarkonium dynamics.

Although it is no doubt that the open system approach is the most consistent quantum mechanical treatment of in-medium quarkonium, there exist more advantageous descriptions in some circumstances.
When more than a few heavy quark pairs are produced in initial hard processes of nuclear collisions, recombination of initially unpaired heavy quarks into a quarkonium at the freezeout is not negligible.
In such a case, open system must contain all the heavy quarks and the practical calculation of master equation is not feasible and instead rate equation is often used to describe the equilibration process of quarkonium.
Analysis by the rate equation \cite{Zhao:2011cv, Song:2011xi, Zhou:2014kka, Du:2022uvj} pointed out that the recombination is indeed an important mechanism for charmonium ($J/\psi$) production at the LHC energy \cite{Abelev:2013ila, Adam:2016rdg}, which makes the open system descriptions not very useful in this setup.
Therefore, our main target will be bottomonium, which is a rare and cleaner probe and allows us to focus on the individual pair produced in an initial hard scattering.

The purpose of this paper is to show the numerical simulations of the Lindblad master equations for quarkonium in the QGP (preliminary results are reported in \cite{Akamatsu:2021dot}) and to analyze them to gain simpler physical descriptions by means of dipole approximation.
It turns out that the steady-state of the Lindblad equation is the Boltzmann distribution within statistical errors.
The equilibration is confirmed for the first time in the non-Abelian case with color-singlet and octet channels and is made possible by the inclusion of environmental memory effects, or the heavy quark recoil, in the Lindblad equation.
The same conclusion has been drawn for the Abelian case \cite{Miura:2019ssi}.
Given that knowledge about the steady-state of the Lindblad equation is limited \cite{rivas2012open}, e.g., \cite{spohn1977algebraic, schirmer2010stabilizing}, it is of vital importance to check its thermalization property.
We further study how the color-singlet and the octet sectors interplay with each other during the equilibration process by calculating the density matrix evolution.
We compare this equilibration process with a simpler Lindblad simulation where quarkonium is approximated as a dipole \cite{Brambilla:2016wgg, Brambilla:2017zei, Akamatsu:2020ypb}.
The dipole approximation is found to be quantitatively useful for a short but long enough time for the relativistic heavy-ion collisions.
This supports the application of such Lindblad equation to the phenomenological studies of quarkonium evolution as is done in \cite{Brambilla:2020qwo, Brambilla:2021wkt, Brambilla:2022ynh}, which is practically beneficial because the numerical cost is much less.

This paper is organized as follows.
In section \ref{sec:QBM}, we first introduce two basic regimes of open quantum systems, i.e. the quantum Brownian and the quantum optical regimes.
We then present the Lindblad equation for the quarkonium is presented in the quantum Brownian regime.
In section \ref{sec:Numerics}, we show our results of numerical simulation.
In particular, we focus on the equilibration process, density matrix evolution, and the validity of the dipole approximation.
In section \ref{sec:Conclusion}, we summarize our paper.

\section{Quantum Brownian motion of quarkonium}\label{sec:QBM}
\subsection{Two regimes of open quantum systems}
Let us start from a general remark on the regimes of open quantum systems \cite{BRE02, Akamatsu:2020ypb}.
This is very important because the useful basis of open system depends on the regime.
In the open quantum system, there are three important time scales: environmental correlation time ($\tau_E$), system relaxation time ($\tau_R$), and system proper timescale ($\tau_S$).
To derive Markovian master equation when the system-environment coupling is weak, it is necessary to assume hierarchy of time scales.

If $\tau_E\ll \tau_R, \tau_S$ is satisfied, the system in the quantum Brownian regime \cite{Caldeira:1982iu}.
In this regime, the fast environmental modes couple to an almost ``frozen" system.
For example, if the system-environment coupling is given by $H_\text{sys-env}\sim X(t)x(t)$, where $X(t)$ and $x(t)$ are system and environmental operators, the system is perturbed by the environment approximately through $X(t)$, because the system evolution is slow $X(t-\tau_E)\simeq X(t)-{\tau_E}\dot X(t) \simeq X(t)$\footnote{
We adopt the interaction picture for $H_\text{sys-env}$.}.
This approximation scheme is the {\it derivative expansion}, which is frequently used in many other fields.

If $\tau_E, \tau_S \ll \tau_R$ is satisfied, the system is in the quantum optical regime.
In this regime, the above approximation is not always possible\footnote{
It can be possible when $\tau_E \ll \tau_S \ll \tau_R$.}.
Instead, the system energy basis $|\epsilon\rangle$ provides a good description.
In this basis, $X(t)=\sum e^{-i\omega t}X_{\epsilon,\epsilon'}|\epsilon\rangle\langle\epsilon'|$ contains rapid phases $\omega\equiv \epsilon'-\epsilon$.
If transition spectrum is discrete with $\Delta\omega = |\omega-\omega'|\sim 1/\tau_S$, interference between transitions with different energy gaps $\omega\neq \omega'$ is effectively lost at time scale of $\tau_R$ since $e^{-i\Delta\omega\tau_R}\sim e^{-i\tau_R/\tau_S}\sim 0$.
This approximation scheme is called {\it rotating wave approximation}.

In the quantum Brownian regime, the density matrix with canonical variables $\langle X|\rho|X'\rangle$ is reasonably convenient while in the quantum optical regime, the density matrix with eigenstate basis $\langle\epsilon|\rho|\epsilon'\rangle$ is preferred.
Below, we analyze the quantum Brownian motion of a heavy quark pair, whose master equation is solved in the position representation.
Note that descriptions based on the bound state levels are inconvenient, if not impossible, because typical quantum states are superposition of many levels.
The condition for quantum Brownian motion $\tau_E \ll \tau_S, \tau_R$ in the case of quarkonium will be examined in the next section\footnote{
If $\tau_E \sim \tau_S \ll \tau_R$, one would expect that quarkonium is in the quantum optical regime.
However, one cannot use the rotating wave approximation because the transition spectrum takes continuous values above the threshold $|\omega|>E_b$ and $\Delta \omega=0$.
One needs to take a classical limit and to treat each of the quarkonium bound states as point-like molecules, by which coupled Boltzmann equations for quarkonium and heavy quarks are obtained \cite{Yao:2018nmy, Akamatsu:2020ypb}.
}.

\subsection{Lindblad equation for quarkonium}
The dynamics of quarkonium in the QGP in the quantum Brownian regime is modeled by the following Lindblad equation
\begin{align}
\label{eq:Lindblad}
&\frac{d}{dt}\rho(t) = -i\left[\frac{p^2}{M}  + \Delta H, \rho\right]\\
& \quad +\sum_{n}\int\frac{d^3k}{(2\pi)^3} \left[
C_n(\vec k)\rho C_n^{\dagger}(\vec k)
-\frac{1}{2}\left\{
C_n^{\dagger}(\vec k) C_n(\vec k), \rho
\right\}
\right]. \nonumber
\end{align}
Here, $\rho(t)$ is the reduced density matrix for the {\it relative} motion of the quarkonium, which consists of color-singlet and octet internal states $\rho(t)=\rho_s|s\rangle\langle s| + \rho_o|o\rangle\langle o|$.
The effect of quantum and thermal fluctuations of gluons is given in $\Delta H$ and dissipator part containing $C_n(\vec k) \ (n=+,-,d,f)$.
They are parameterized by two functions $U(r)$ and $\gamma(r)$ (or its Fourier transform $\gamma(k)$),
\begin{subequations}
\label{eq:Lindblad_HC}
\begin{align}
&\Delta H= \left[U(r) - \frac{\left\{\vec p, \vec\nabla \gamma(x) \right\}}{4MT} \right]
\left(\frac{4}{3}|s\rangle\langle s| - \frac{|o\rangle\langle o|}{6}\right), \\
&C_+(\vec k) = \sqrt{\frac{4\gamma(k)}{3}}V_+(\vec k)|o\rangle\langle s|, \\
&C_-(\vec k) = \sqrt{\frac{\gamma(k)}{6}}V_-(\vec k) |s\rangle\langle o|,\\
&C_{d}(\vec k) =\sqrt{\frac{5\gamma(k)}{12}}V_{d}(\vec k) |o\rangle\langle o|,\\
&C_{f}(\vec k) =\sqrt{\frac{3\gamma(k)}{4}}V_{f}(\vec k) |o\rangle\langle o|,
\end{align}
\end{subequations}
where $T$ is the temperature of QGP, and $M$ is the heavy quark kinetic mass.
The temperature is assumed to be $T\ll M$ so that thermal pair-creation of heavy quarks can be neglected. 

The Hamiltonian shift $\Delta H$ represents the real part of system self-energy caused by the coupling to the environment\footnote{
From the structure of the Lindblad equation \eqref{eq:Lindblad}, the self-energy is readily extracted as
\begin{align}
\Delta H  - \frac{i}{2}\sum_{n}\int\frac{d^3k}{(2\pi)^3} 
C_n^{\dagger}(\vec k) C_n(\vec k).
\end{align}
Conversely, by examining the imaginary part of the self-energy diagrams, one can infer the physical process described by the Lindblad operators.
}.
$U(r)$ is the potential term given by virtual exchange of environmental gluons\footnote{
In the Potential Non-Relativistic QCD (pNRQCD) effective theory, $U(r)$ is interpreted as a part of the system Hamiltonian because gluon with momentum $k\sim 1/r$ is already integrated out in the vacuum before coupling the system and the finite temperature environment.
}.
The next term corrects the potential term, which is obtained in the limit of static heavy quark, by introducing the effect of heavy quark recoils.
Similar term ($\propto \{\vec p,\vec x\}/MT$) is also found in the Caldeira-Leggett model \cite{Caldeira:1982iu} for the quantum Brownian motion.

The Lindblad operators $C_n(\vec k)$ describe scatterings with momentum transfer $\vec k$ in various color channels in the singlet $|s\rangle$ and the octet $|o\rangle$, taking place with rates such as $4\gamma(k)/3$ for $n=+$.
The subscripts $d$ and $f$ indicate structure constants $d_{abc}$ and $f_{abc}$ of SU(3) algebra involved in a scattering process.
Schematically, in one-gluon ($g^*$) exchange process, an octet quarkonium scatters into
\begin{align}
g^*(a)|c\rangle \sim \sum_b\left(C_d d_{abc} + C_f if_{abc} \right)|b\rangle + C_-\delta_{ac}|s\rangle,
\end{align}
where $a,b,c$ are the octet charges.
Note that $|o\rangle$ above denotes the octet after projection while $|b\rangle$ and $|c\rangle$ are individual octet states.
See \cite{Akamatsu:2020ypb} for more technical details.
The operators $V_n(\vec k)\simeq e^{i\vec k\cdot \vec r/2} \pm e^{-i\vec k\cdot \vec r/2} +\cdots$ describe heavy quark momentum shifts in a scattering\footnote{
Recall that $\vec r=\vec r_Q-\vec r_{Q_c}$ is the relative coordinate between the heavy quark pair.
Then, a scattering $\vec p_Q\to \vec p_Q + \vec k$ shifts the relative momentum $\vec p=(\vec p_Q-\vec p_{Q_c})/2\to \vec p+\vec k/2$ and $\vec p_{Q_c}\to \vec p_{Q_c} + \vec k$ shifts $\vec p\to\vec p - \vec k/2$.
The leading term $e^{i\vec k\cdot \vec r/2} \pm e^{-i\vec k\cdot \vec r/2}$ represents the superposition of these two scatterings with relative signs (or phases in general) depending on the color channels.
}, where the omitted terms take account of the heavy quark recoils.
Their explicit forms are
\begin{subequations}
\label{eq:Lindblad_V}
\begin{align}
V_{\pm}(\vec k) &=
\left(1-\frac{\vec k\cdot\vec p}{4MT} \pm \frac{3U(r)}{8T}\right)e^{i\vec k\cdot\vec r/2}
-[\vec k\to-\vec k],\\
V_{d/f}(\vec k) &=
\left(1-\frac{\vec k\cdot\vec p}{4MT}\right)e^{i\vec k\cdot\vec r/2} 
\mp [\vec k\to-\vec k].
\end{align}
\end{subequations}
The terms due to the heavy quark recoils $\propto 1/T$ are required by the fluctuation-dissipation relation of thermal correlation functions in the QGP and are responsible for directing the quarkonium system toward the equilibrium \cite{Akamatsu:2014qsa, Akamatsu:2020ypb, Akamatsu:2021dot}.
This Lindblad equation describes the decoherence phenomena and dissipation of quarkonium.

\begin{figure}[!t]
\begin{center}
\includegraphics[width=0.5\textwidth, angle=0]{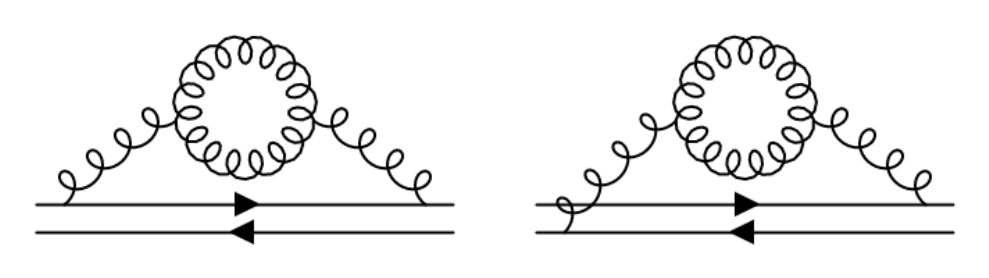}
\end{center}
\caption{Examples of the self-energy diagrams.
The directed lines are heavy quark and antiquark propagators and the curly lines are gluon propagators.
There are three ways of attaching the gluon propagator to the heavy (anti)quark propagators, of which two are shown.
The light quarks can also run in the internal loops.
The loops are calculated in the hard-themal loop approximation and are resummed in the gluon propagators, which is not explicitly shown.
}
\label{fig:diagrams_NR}
\end{figure}

Original derivation \cite{Akamatsu:2014qsa, Akamatsu:2020ypb} was performed in the weak-coupling expansion $g\ll 1$ of Non-Relativistic QCD (NRQCD) and by assuming a particular configuration of heavy quark pair $r\sim 1/gT$.
Self-energy diagrams considered in the derivation are shown in the Fig.~\ref{fig:diagrams_NR}.
The three time scales are estimated as
\begin{subequations}
\begin{align}
&\tau_E\sim 1/gT &\text{(color electric scale)}, \\
&\tau_S \sim 4\pi/g^3T  &\text{(inverse potential energy)},\\
&\tau_R \sim M/g^4T^2 &\text{(kinetic equilibration \cite{Moore:2004tg})}.
\end{align}
\end{subequations}
Therefore, the condition $\tau_E\ll\tau_R,\tau_S$ reads
\begin{align}
g \ll (4\pi)^{1/2}, (M/T)^{1/3}.
\end{align}
The weak coupling assumption $g\ll 1$ is thus enough to justify the quantum Brownian motion.
In this case, the dynamics of quarkonium is in the quantum Brownian regime and the functions $U(r)$ and $\gamma(k)$ are
\begin{align}
\label{eq:Lindblad_UG}
U(r) = -\frac{g^2}{4\pi r}e^{-m_Dr}, \quad
\gamma(k) = g^2 T\frac{\pi m_D^2}{k(k^2+m_D^2)^2}.
\end{align}
Here $m_D = gT\sqrt{1+N_f/6}$ is the Debye screening mass for $N_f$ massless quark flavors.
To be precise, the center-of-mass and relative dynamics are coupled and the former cannot be traced out.
By projecting its full dynamics on the fixed center-of-mass momentum $\vec P=\vec 0$, one gets the above Lindblad equation for the relative dynamics \cite{Akamatsu:2018xim}.

Although this Lindblad equation is obtained with a rather limited condition, there is a reason to believe that it captures essential features of quarkonium dynamics in the QGP.
If the heavy quark pair distance is small, one can take a short distance {\it extrapolation} of \eqref{eq:Lindblad} by a relatively mild assumption for the (non-)analyticity at the origin
\begin{subequations}
\begin{align}
\label{eq:U_smallr}
U(r) &= -\frac{\alpha}{r} + U_0 + \frac{1}{2}U_2 r^2 + \cdots,\\
\label{eq:G_smallr}
\gamma(r) &= \gamma_0 + \frac{1}{2}\gamma_2 r^2 + \cdots, \quad \gamma_2 < 0,
\end{align}
\end{subequations}
to get (by replacing $\sum_n\int\frac{d^3k}{(2\pi)^3}$ with $\sum_{n}\sum_{j=x,y,z}$ in \eqref{eq:Lindblad})
\begin{subequations}
\label{eq:Lindblad_dipole_HC}
\begin{align}
\Delta H &= \left[-\frac{\alpha}{r} + \frac{1}{2}U_2 r^2 \right]
\left(\frac{4}{3}|s\rangle\langle s| - \frac{|o\rangle\langle o|}{6}\right)\nonumber \\
& \quad - \frac{\gamma_2}{4MT}\{\vec r, \vec p\}
\left(\frac{4}{3}|s\rangle\langle s| + \frac{7}{12}|o\rangle\langle o|\right), \\
C_{+j} &= \sqrt{\frac{-4\gamma_2}{3}}\left[r_j - \frac{3\alpha r_j}{8Tr} + \frac{ip_j}{2MT}\right]|o\rangle\langle s|, \\
C_{-j} &= \sqrt{\frac{-\gamma_2}{6}}\left[r_j + \frac{3\alpha r_j}{8Tr} + \frac{ip_j}{2MT}\right] |s\rangle\langle o|,\\
C_{dj} &=\sqrt{\frac{-5\gamma_2}{12}}\left[r_j + \frac{ip_j}{2MT}\right] |o\rangle\langle o|.
\end{align}
\end{subequations}
See \cite{Akamatsu:2020ypb} for more details.
This Lindblad form is almost identical to that obtained from the pNRQCD effective field theory, which is based on the multi-pole expansion in terms of heavy quark pair distance \cite{Brambilla:2016wgg, Brambilla:2017zei, Akamatsu:2020ypb}\footnote{
There is a minor technical difference.
To get the Lindblad equation at $\mathcal O(r^2)$, one needs to start from effective Lagrangian expanded up to $\mathcal O(r^2)$.
The Lindblad equation from pNRQCD is obtained from the Lagrangian up to $\mathcal O(r)$. 
}.

\begin{figure}[!t]
\begin{center}
\includegraphics[width=0.3\textwidth, angle=0]{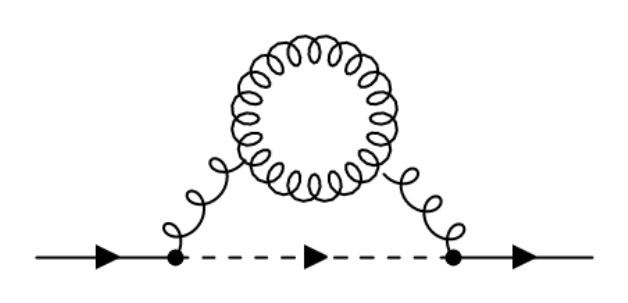}\\
\includegraphics[width=0.3\textwidth, angle=0]{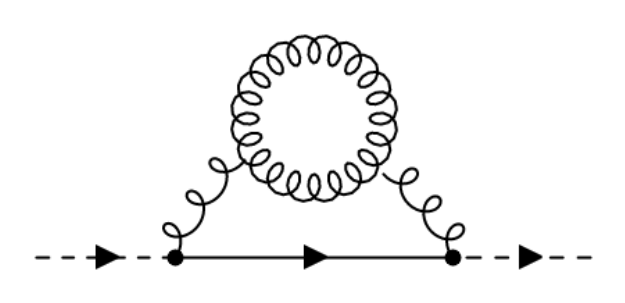}\\
\includegraphics[width=0.3\textwidth, angle=0]{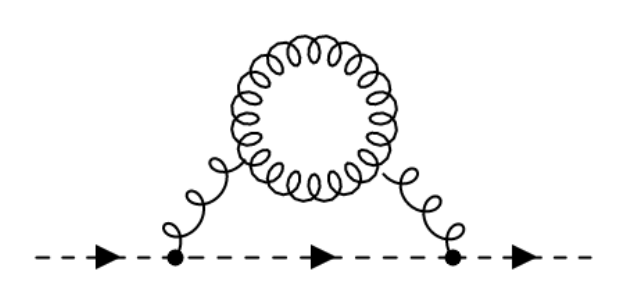}\\
\end{center}
\caption{Examples of the self-energy diagrams.
The directed lines are singlet quarkonium propagators, the dashed directed lines are octet quarkonium propagators, and the curly lines are gluon propagators.
From the top to the bottom, the diagrams depict self-energy contributions from singlet-to-octet, octet-to-singlet, and octet-to-octet virtual processes.
The gluon propagators with one-loop correction are drawn just for concreteness.
In the actual calculation, fully nonperturbative and gauge invariant two-point function of color electric fields are attached to the quarkonium propagators with vertices $\propto gr$.
To ensure the gauge invariance, the fundamental Wilson lines are attached between these color electric fields.
}
\label{fig:diagrams_pNR}
\end{figure}

The derivation from pNRQCD does not need to assume the weak coupling $g\ll 1$ as long as the multi-pole expansion can be justified, i.e. $r\ll T^{-1}, m_D^{-1}$.
Self-energy diagrams considered in the derivation are shown in the Fig.~\ref{fig:diagrams_pNR}.
The three time scales in the non-perturbative region can be estimated by
\begin{subequations}
\begin{align}
&\tau_E\sim 1/m_D\sim 1/\pi T &\text{(typical thermal scale)}, \\
&\tau_S\sim 4/[M(4\alpha /3)^2] &\text{(inverse binding energy)},\\
&\tau_R\sim M/T^2 & \text{(kinetic equilibration)}.
\end{align}
\end{subequations}
For bottomonium, $M\approx 4.8 {\rm GeV}$ and $4\alpha/3\sim 0.3-0.4$.
Then, one of the conditions of quantum Brownian regime $\tau_E\ll \tau_S$ reads
\begin{align}
\pi T \gg (0.11-0.19) {\rm GeV},
\end{align}
which covers a large portion of temperature region available in the heavy-ion collisions, roughly $0.1{\rm GeV} \lesssim T \lesssim 0.5{\rm GeV}$.
The other one $\tau_E\ll\tau_R$ is safely satisfied for bottom quarks in the heavy-ion collisions.
In the Lindblad equations from pNRQCD, the coefficients $U_2$ and $\gamma_2$ are replaced with physical quantities with gauge invariant and non-perturbative definitions.
To be specific, $U_2$ and $\gamma_2$ are responsible for the quarkonium mass shift and width at finite temperature and $\frac{4}{3}\gamma_2$ is the heavy quark momentum diffusion constant.

In this way, we expect that functions $U(r)$ and $\gamma(r)$ do exist which are compatible with both the multi-pole expansion $r\ll T^{-1}, m_D^{-1}$ and the weak-coupling expansion $g\ll 1$ in soft regime $r\sim 1/gT$.

\section{Numerical simulation}\label{sec:Numerics}
\subsection{Setup}
In this paper, we show the results of numerical simulations of the Lindblad model \eqref{eq:Lindblad} with
\begin{subequations}
\label{eq:UG_model}
\begin{align}
U(r)&=-\frac{\alpha}{r}e^{-m_Dr},\quad
\gamma(r) = \gamma_0 e^{-r^2/\ell_{\rm corr}^2}, \\
\label{eq:UG_model_parameters}
\alpha&=0.225, \quad m_D =2\ell_{\rm corr}^{-1}=2T , \quad \gamma_{0}= \frac{T}{\pi}.
\end{align}
\end{subequations}
The functions and its parameters are borrowed from our previous work \cite{Miura:2019ssi}, which are motivated from various known results about the heavy quark complex potential.
\begin{itemize}
\item $U(r)$ of \eqref{eq:UG_model} models the short-range Coulomb attraction and the long-range screening.
\begin{itemize}
\item $\frac{4}{3}\alpha=0.3$ is chosen from the Coulomb part of the Cornell potential \cite{Bali:2000gf, Alford:2013jva}.
\item $m_D=2T$ is obtained by substituting $g=2, N_f=0$ into its perturbative result.
\end{itemize}
\item $\gamma(r)$ of \eqref{eq:UG_model} models the essential behaviors of its perturbative result (Fourier transform of \eqref{eq:Lindblad_UG}), which is a monotonically decreasing function with typical length scale $1/m_D$, decaying to 0 at long distance $r\gg 1/m_D$.
\begin{itemize}
\item The relation $m_D=2\ell_{\rm corr}^{-1}$ is obtained by equating the full width half maximum of $\gamma(r)$ in \eqref{eq:UG_model} and that from \eqref{eq:Lindblad_UG}.
\item $\gamma_0=\gamma(r=0)$ is obtained by substituting $g=2$ into its perturbative result $g^2 T/4\pi$.
\end{itemize}
\end{itemize}
Note that the short distance behavior of $U(r)$ in \eqref{eq:UG_model} takes the form of \eqref{eq:U_smallr} only up to $U_0$, but the difference is subleading in $r\to 0$.

The simulation is mostly done with $T=0.1 M$, which is $T\approx 0.48 {\rm GeV}$ for bottomonium.
The simulation is performed by the Quantum State Diffusion (QSD) method \cite{gisin1992quantum}, which gives an algorithm to sample random wave functions evolving according to the nonlinear stochastic Schr\"odinger equation.
Its stochastic evolution equation can be written in terms of the Lindblad operators as summarized in the Appendix \ref{app:QSD}.
In our simulation, we sampled thousands of random wave functions.
For specific forms of the QSD equations, see \cite{MiuraThesisOsakaU}.
The numerical cost to evolve one sample increases with the square of the lattice points $\sim N_x^2$ because the number of the Lindblad operators is $\sim N_x$.
On the contrary, the dipole approximation reduces the number of the Lindblad operators to $\mathcal O(1)$ and the numerical cost increases only linearly with $N_x$.

\begin{figure}[!t]
\begin{center}
\includegraphics[height=0.4\textwidth, angle=-90]{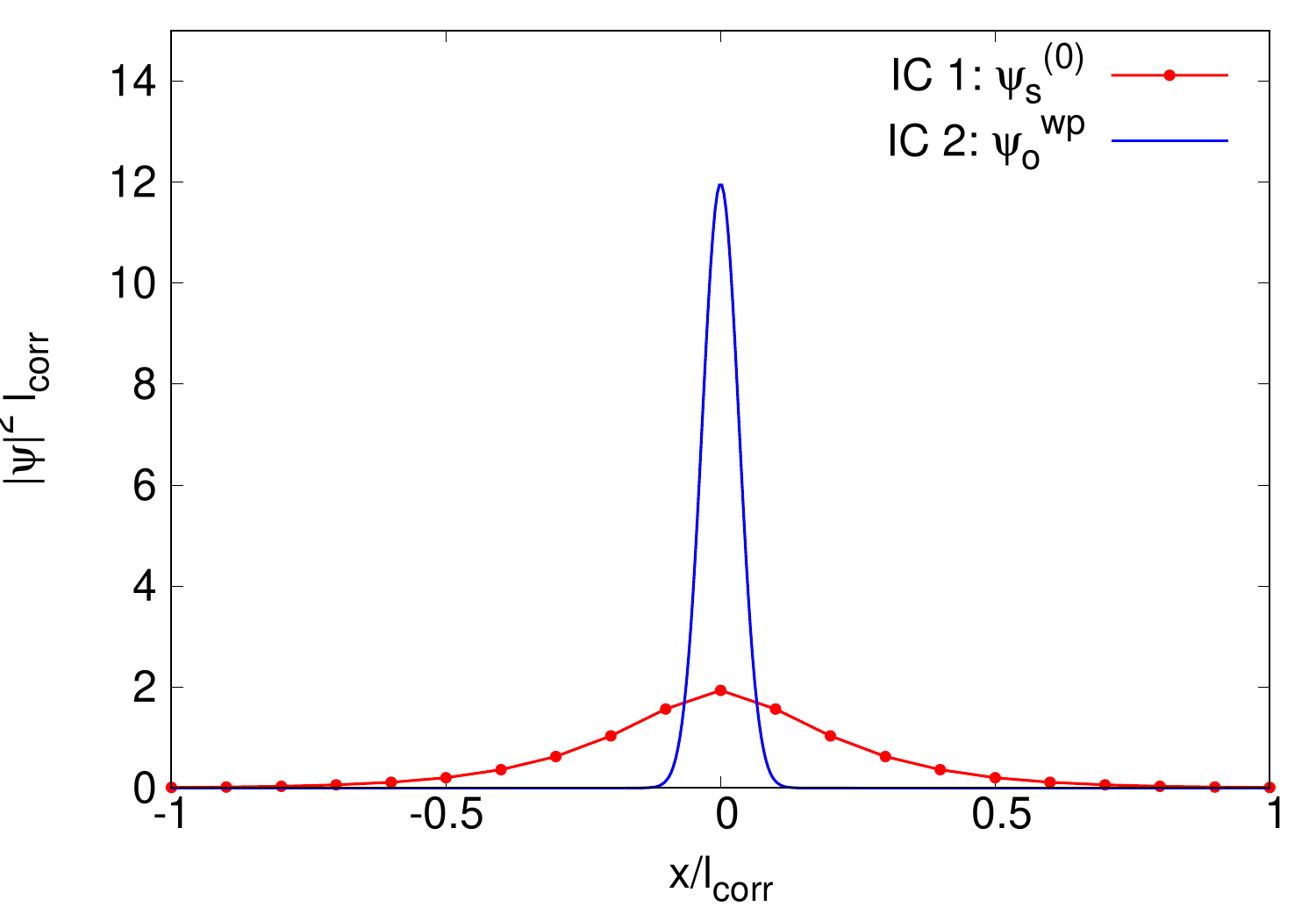}
\end{center}
\caption{Spatial profile of initial wave functions for IC~1 and IC~2.
They are both localized compared to $\ell_{\rm corr}$, which is a {\it color resolution scale} of the QGP medium.
}
\label{fig:initial}
\end{figure}

The system is prepared on a one-dimensional lattice with $N_x=255$ points with $\Delta x=1/M$ and the wave functions satisfy the periodic boundary condition.
The origin $x=0$ is located at the center (the 128th lattice point) and the singular behavior of $U(r)\sim -\alpha/r$ is tamed by a cutoff $1/|x|\to 1/\sqrt{x^2+x_c^2}\equiv 1/|x|_{\rm reg}$ with $x_c=1/M$.
Two different initial conditions (ICs) are adopted,
\begin{subequations}
\begin{align}
\text{IC~1}&: \ \psi_s(x)=\psi_s^{(0)}(x), \quad \psi_o(x)=0,\\
\text{IC~2}&: \ \psi_s(x)=0, \quad \psi_o(x) = \frac{e^{-x^2/2a^2}}{(\pi a^2)^{1/4}}\equiv \psi_o^{\rm wp}(x),
\end{align}
\end{subequations}
where $\psi_s^{(i)} \ (i=0,1,2,\cdots)$ denotes the singlet ground state ($i=0$) or an excited state ($i\geq 1$) in the screened potential \eqref{eq:UG_model}.
Note that only the ground state is bound in our screened potential in one dimension.
The size of the wave packet $\psi_o^{\rm wp}(x)$ is $a = \frac{0.2}{\sqrt{2}} \cdot \frac{1}{(4/3)M\alpha}$.
The wave functions are plotted in the Fig.~\ref{fig:initial}.
These initial conditions are motivated by two limiting scenarios for the initial conditions of the quarkonium in heavy-ion collisions --- the bound state formation time is short enough (IC~1) and is too long (IC~2) compared to the initialization time.
The IC~2 is similar to some of the initial conditions used in \cite{Casalderrey-Solana:2012yfo, Brambilla:2016wgg, Brambilla:2017zei} and is based on the fact that the heavy quark pair production takes place locally by initial hard processes and mainly in the octet channel \cite{Cho:1995vh, Cho:1995ce}.

\begin{figure}[!t]
\begin{center}
\includegraphics[width=0.4\textwidth]{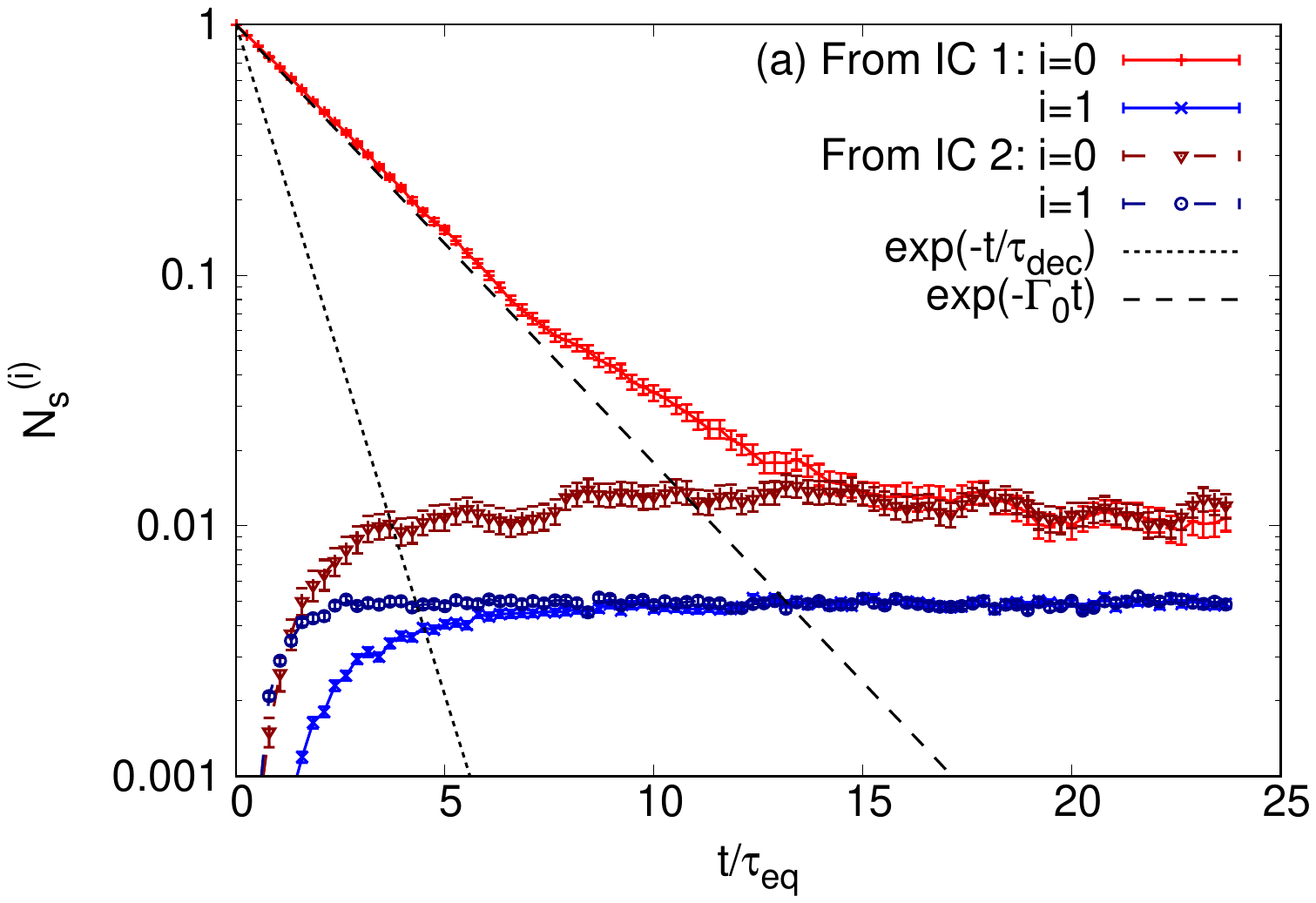}
\includegraphics[width=0.4\textwidth]{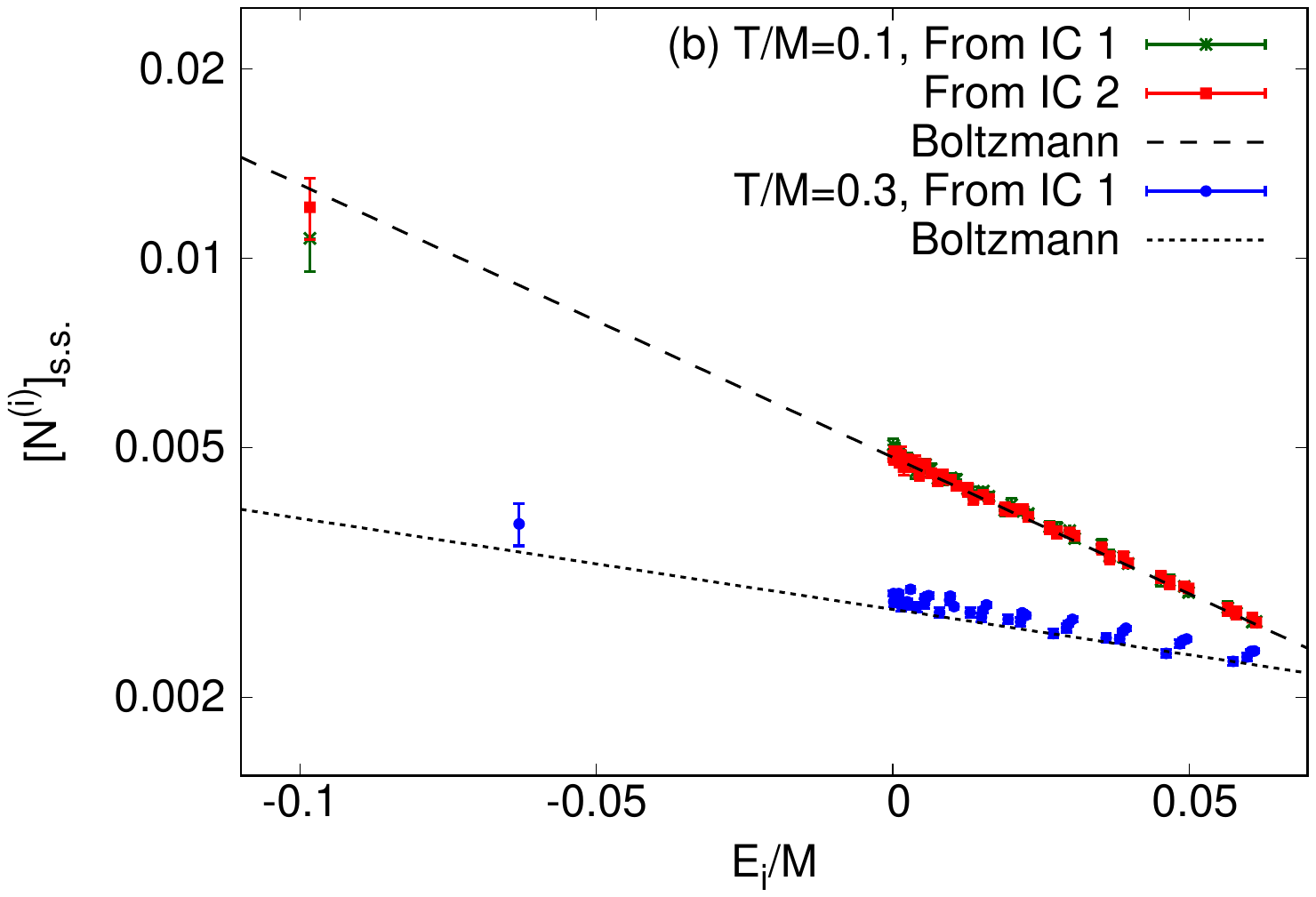}
\end{center}
\caption{Equilibration process of quarkonium states.
(a) Time evolutions from ICs 1 and 2 are plotted for the occupations of the lowest two levels in the singlet channel.
(b) Steady-state occupations for the low-lying modes for both singlet and octet are plotted.
For both $T=0.1M$ (from ICs 1 and 2) and $T=0.3M$ (from IC~1), the distribution in the steady state is well approximated by the Boltzmann distribution $A\cdot e^{-E_i/T}$ with corresponding temperatures, whose normalization, $A=0.00482$ and $0.00276$ for $T=0.1M$ and $0.3M$, respectively, is calculated by the complete spectrum of quarkonium states.
}
\label{fig:equilibration}
\end{figure}

\subsection{Equilibration of the system}
In Fig.~\ref{fig:equilibration}(a), we show the evolution of state occupation numbers $N_{s}^{(i)}(t)\equiv \langle\psi_s^{(i)}|\rho(t)|\psi_s^{(i)}\rangle$ for $i=0,1$ starting from IC~1 and IC~2.
Damping time due to friction $\tau_{\rm eq}\simeq 236/M$ and quantum decoherence time $\tau_{\rm dec} \simeq 191/M$ for the ground state will serve as useful reference scales.
They are defined as
\begin{subequations}
\begin{align}
\frac{1}{\tau_{\rm eq}}&\equiv -\frac{\langle \dot p\rangle}{\langle p\rangle}\Bigr|_{U(r)=0}=\frac{4\gamma_0}{3MT\ell_{\rm corr}^2} ,\\
\frac{1}{\tau_{\rm dec}}&\equiv -\dot N_s^{(0)}(0)\Bigr|_{\rm recoilless} =\frac{8}{3}\int dx (\gamma(0)-\gamma(x))|\psi_s^{(0)}(x)|^2,
\end{align}
\end{subequations}
and computed at $T=0.1M$.
For a wave function with macroscopic size, these time scales are hierarchical $\tau_{\rm eq}\gg \tau_{\rm dec}$.
However, the bound state wave function $\psi_s^{(0)}(x)$ is localized with $\sqrt{\langle x^2\rangle }\simeq 2.67/M \ll 10/M=\ell_{\rm corr}$ and $\tau_{\rm eq}$ is not much different from $\tau_{\rm dec}$.

From the figure, one can observe that the relaxation of the excited singlet state takes place on the time scale of $\tau_{\rm eq}$ in both initial conditions.
The relaxation takes place not only in the phase space but also in the color space.
From IC~1, the singlet ground state first gets excited to the octet and then returns to the singlet.
From IC~2, the initial state is already in the octet and thus the equilibration process is faster than from IC~1. 

As for the singlet ground state, the relaxation from IC~2 takes place roughly in the time scale of $\tau_{\rm eq}$.
However, the decay from IC~1 takes about 3 times longer than with $e^{-t/\tau_{\rm dec}}$.
The reason is as follows.
For bound states smaller than $\ell_{\rm corr}$, the decoherence is ineffective, and heavy quark recoils during the decoherence process are no longer subdominant.
The recoil is responsible for irreversible processes such as friction in the classical limit, which prevents dissociation, and thus the decay process takes a longer time.
Indeed, the damping $e^{-\Gamma_0 t}$ with initial decay rate $\Gamma_0$ (defined later in Eq.~\eqref{eq:Gamma_0}) including the recoil effect can quantitatively reproduce the numerical data before thermalization.

In Fig.~\ref{fig:equilibration}(b), the distribution of eigenstates $[N^{(i)}]_{\rm s.s.}$ (both singlet and octet) in the steady state at $t\approx 24\tau_{\rm eq}$ is plotted for $T=0.1M$ and $0.3M$.
Note that the occupation number of an octet eigenstate is defined as $N_{o}^{(i)}(t)\equiv \langle\psi_o^{(i)}|\rho(t)|\psi_o^{(i)}\rangle/8$.
One can see that the distribution can be approximated well by the Boltzmann distribution with the environment temperatures, irrespective of the initial conditions for $T=0.1M$.
The distributions are fitted by a two-parameter function $[N^{(i)}]_{\rm s.s.}=A'\cdot e^{-E_i/T'}$.
The data for IC 1 and 2 with $T=0.1M$ are best fitted by $(A',T'/M)=(0.0049, 0.0994)$ and $(0.0049, 0.1012)$, respectively, and those for $T=0.3M$ are by $(0.0029, 0.2846)$.
The relative errors are $\lesssim 1\%$ except for the last one ($T'/M=0.2846$ with a relative error $\sim 4\%$).
Although the steady state solution of the Lindblad equation \eqref{eq:Lindblad} is not known, the Lindblad operators approximately satisfy the detailed balance relation with the recoil terms and the equilibration to the Boltzmann distribution is expected.

\begin{figure}
\centering
\includegraphics[width=5.5cm,bb=60 0 320 216]{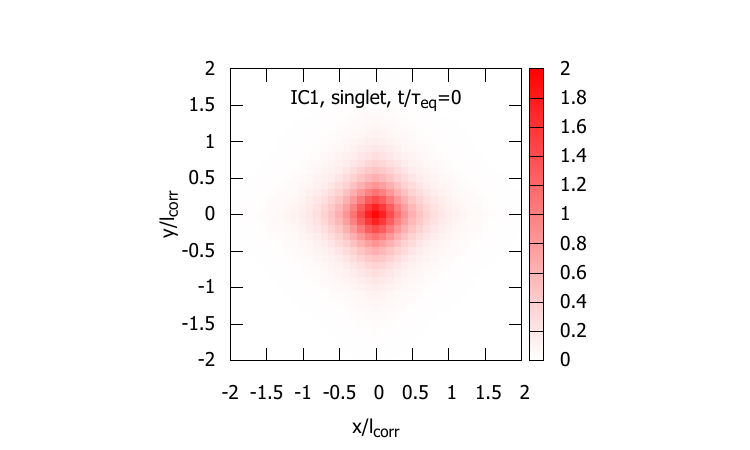}
\includegraphics[width=5.5cm,bb=60 0 320 216]{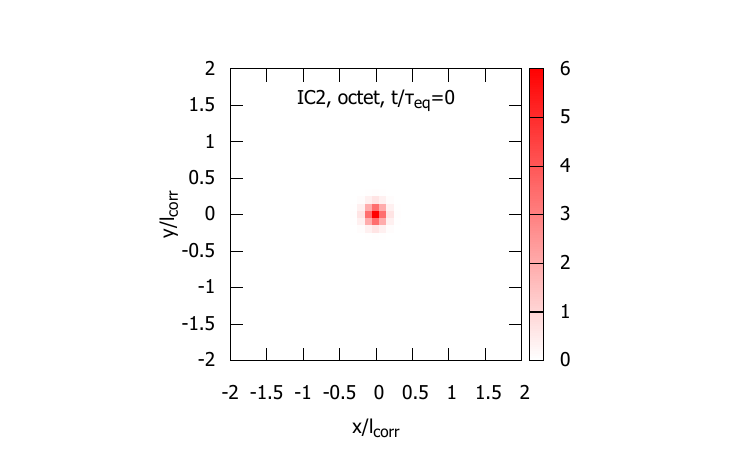}
\caption{Initial density matrices $|\rho_{s}(x,y,0)|\ell_{\rm corr}$ for IC 1 and $|\rho_{o}(x,y,0)|\ell_{\rm corr}$ for IC 2.}
\label{fig:density_matrix_init}
\end{figure}

\subsection{Density matrix}
\begin{figure*}
\centering
\includegraphics[width=5.5cm,bb=60 0 320 216]{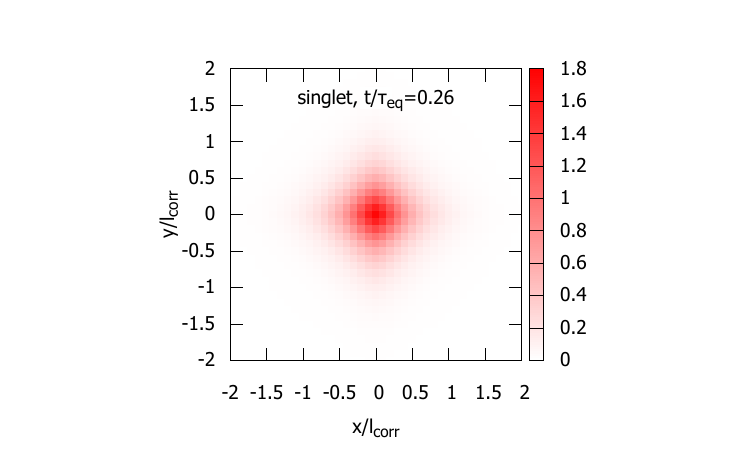}
\includegraphics[width=5.5cm,bb=60 0 320 216]{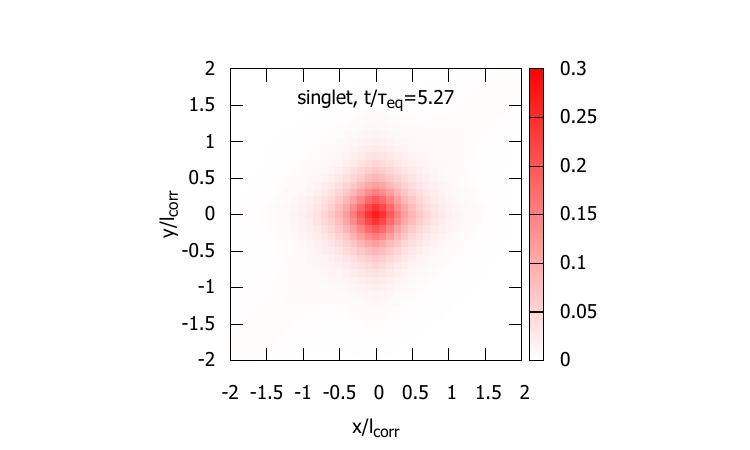}
\includegraphics[width=5.5cm,bb=60 0 320 216]{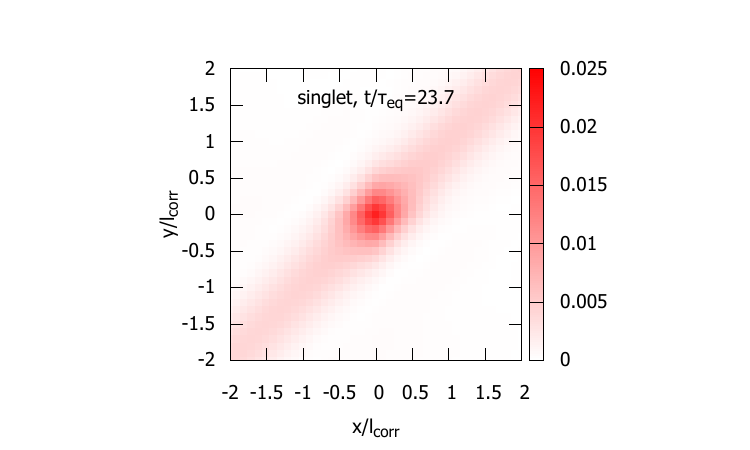}\\
\includegraphics[width=5.5cm,bb=60 0 320 216]{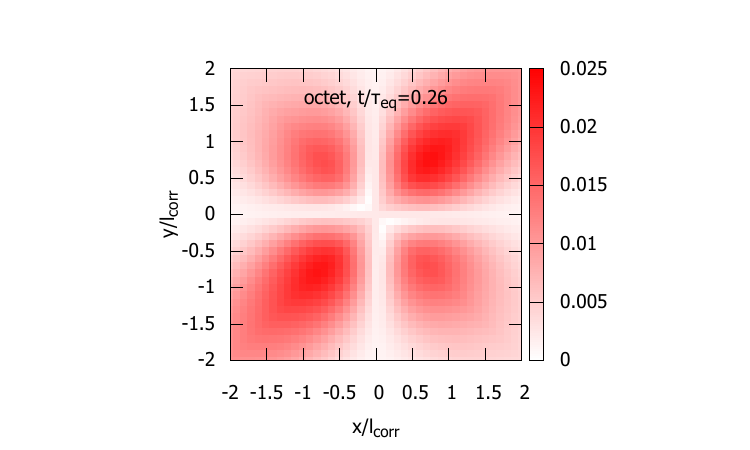}
\includegraphics[width=5.5cm,bb=60 0 320 216]{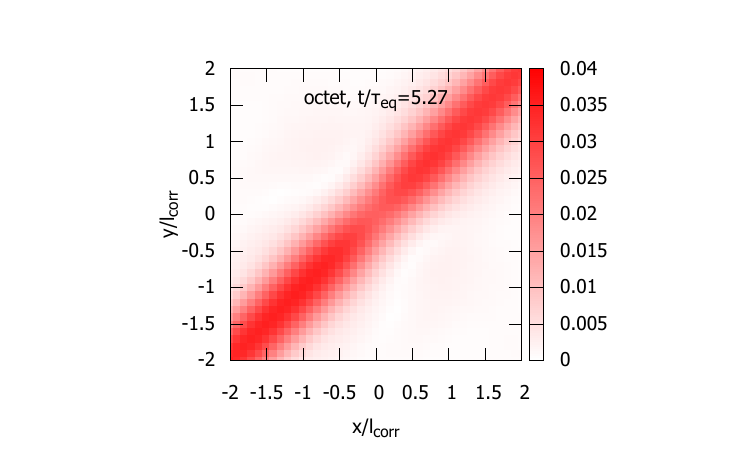}
\includegraphics[width=5.5cm,bb=60 0 320 216]{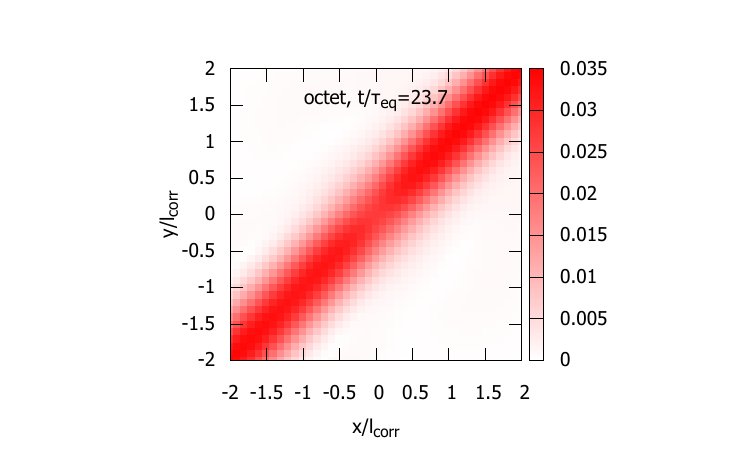}
\caption{Time evolution of the reduced density matrix in the singlet and octet sectors $|\rho_{s,o}(x,y,t)|\ell_{\rm corr}$ at $t=0.26\tau_{\rm eq}, 5.27\tau_{\rm eq}$, and $23.7\tau_{\rm eq}$.
The initial condition is IC 1.
The figure only shows central domain $-2\ell_{\rm corr} \leq x, y \leq 2\ell_{\rm corr}$.
The top figures are the single density matrices and the bottom figures are the octet density matrices.}
\label{fig:density_matrix_IC1}
\end{figure*}

\begin{figure*}
\centering
\includegraphics[width=5.5cm,bb=60 0 320 216]{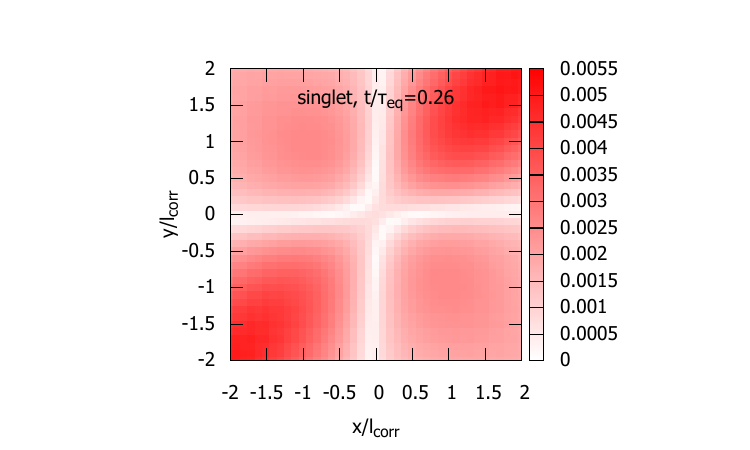}
\includegraphics[width=5.5cm,bb=60 0 320 216]{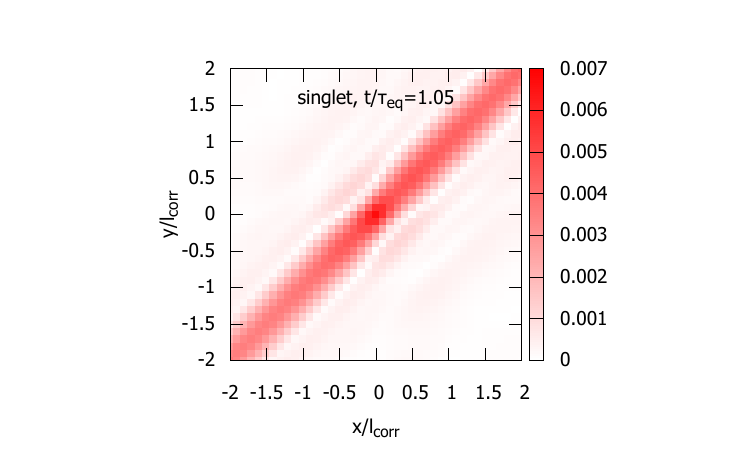}
\includegraphics[width=5.5cm,bb=60 0 320 216]{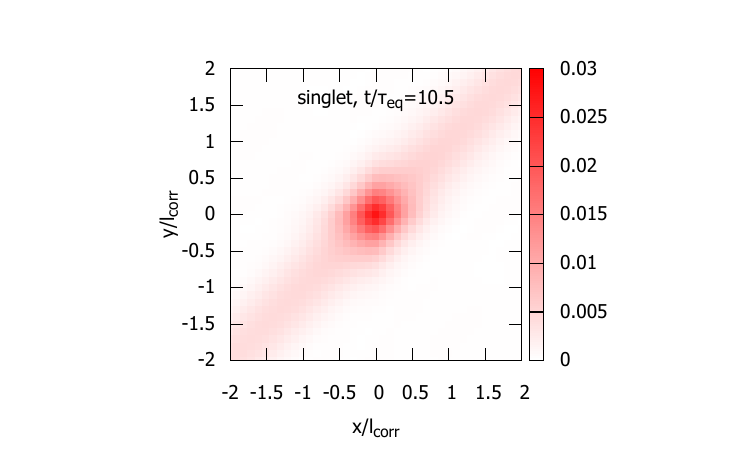}\\
\includegraphics[width=5.5cm,bb=60 0 320 216]{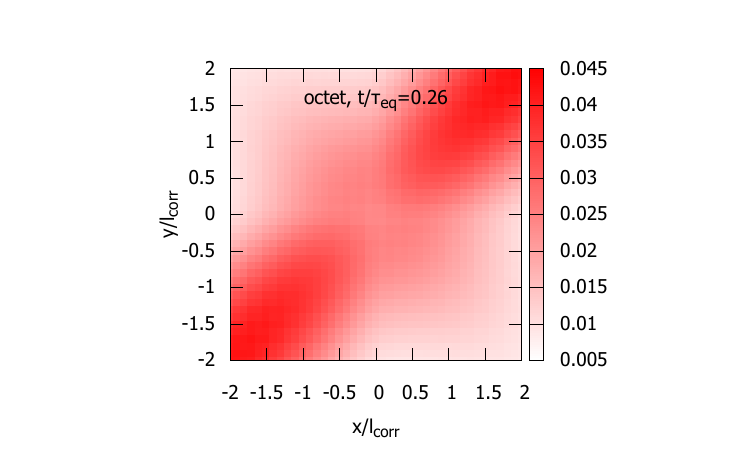}
\includegraphics[width=5.5cm,bb=60 0 320 216]{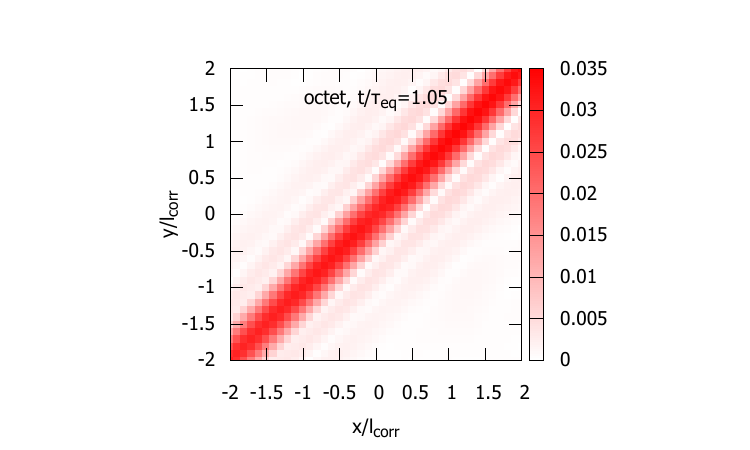}
\includegraphics[width=5.5cm,bb=60 0 320 216]{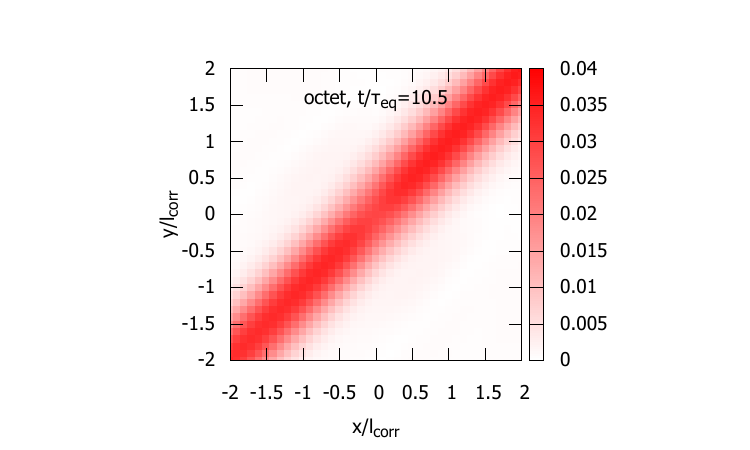}
\caption{Time evolution of the reduced density matrix in the singlet and octet sectors $|\rho_{s,o}(x,y,t)|\ell_{\rm corr}$ at $t=0.26\tau_{\rm eq}, 1.05\tau_{\rm eq}$, and $10.5\tau_{\rm eq}$.
The initial condition is IC 2.
The figure only shows central domain $-2\ell_{\rm corr} \leq x, y \leq 2\ell_{\rm corr}$.
The top figures are the single density matrices and the bottom figures are the octet density matrices.}
\label{fig:density_matrix_IC2}
\end{figure*}

In Figs.~\ref{fig:density_matrix_init}, \ref{fig:density_matrix_IC1}, and \ref{fig:density_matrix_IC2}, time evolution of the density matrix is shown.
The initial density matrices for ICs 1 and 2 are shown in Fig.~\ref{fig:density_matrix_init}.
Both are localized at the origin $x\sim y\sim 0$ within $\lesssim \ell_{\rm corr}$.
In Figs.~\ref{fig:density_matrix_IC1} and \ref{fig:density_matrix_IC2}, the top figures are the singlet density matrices and the bottom figures are the octet density matrices.

From IC 1, the evolution of the density matrix proceeds by the three steps.
\begin{enumerate}
\item[(1)] The singlet ground state is excited to octet as a color dipole.
Note that the singlet-to-octet transition is forbidden at $xy=0$ in the recoilless limit (see Eq.~\eqref{eq:Lindblad_V}) [left; $t=0.26\tau_{\rm eq}$].
\item[(2)] The octet density matrix is diagonalized and gets close to its steady state.
At this stage, the singlet density matrix is still dominated by the ground state and its time evolution is only visible in the magnitude at the origin [center; $t=5.27\tau_{\rm eq}$].
\item[(3)] De-excitation from the octet to the singlet is observed and the system finally reaches its steady state [right; $t=23.7\tau_{\rm eq}$].
\end{enumerate}
From IC 2, the evolution of the density matrix proceeds by the three steps.
\begin{enumerate}
\item[(1)] The octet wave packet expands rapidly and transitions to the singlet state with a wide distribution.
Note that the octet-to-singlet transition is forbidden at $xy=0$ in the recoilless limit  (see Eq.~\eqref{eq:Lindblad_V}) [left; $t=0.26\tau_{\rm eq}$].
\item[(2)] The singlet and octet density matrices are diagonalized.
The octet density matrix is close to its steady state while the singlet density matrix is not because the magnitude at the origin is still small [center; $t=1.05\tau_{\rm eq}$].
\item[(3)] The singlet density matrix now contains enough ground state occupation and the system finally reaches its steady state [right; $t=10.5\tau_{\rm eq}$].
\end{enumerate}

From this simulation, we find that the quarkonium relative distribution extends far beyond the scale $\ell_{\rm corr}$ both for the singlet and the octet.
Naively, one would expect that the dipole approximation does not work.
In the next section, we examine this naive expectation and find that the dipole approximation actually works reasonably well as long as the ground state occupation $N_s^{(0)}(t)$ in the short time scale matters ($t\sim \mathcal O(\tau_{\rm eq})$).

\begin{figure}[!t]
\begin{center}
\includegraphics[width=0.4\textwidth]{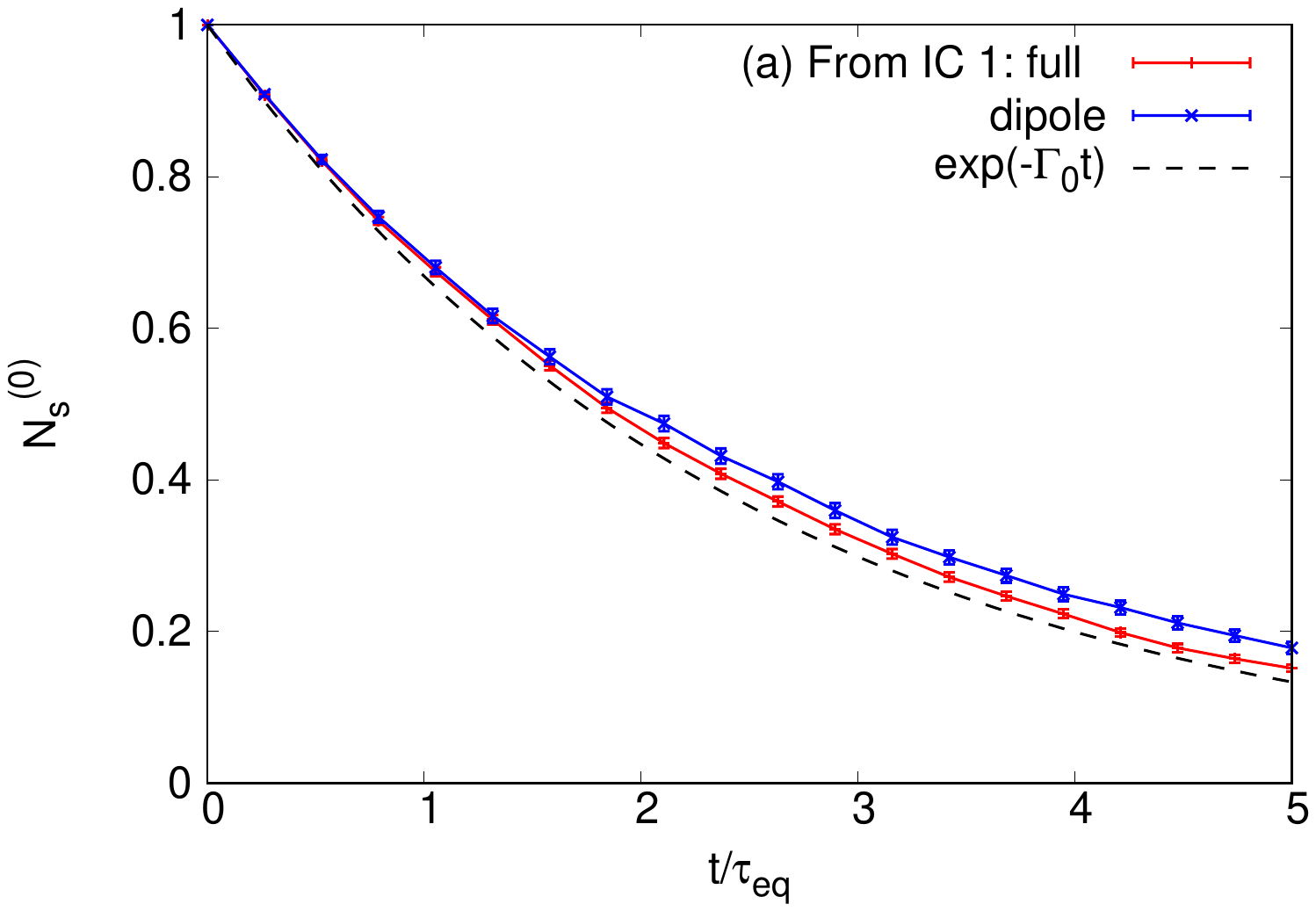}
\includegraphics[width=0.41\textwidth]{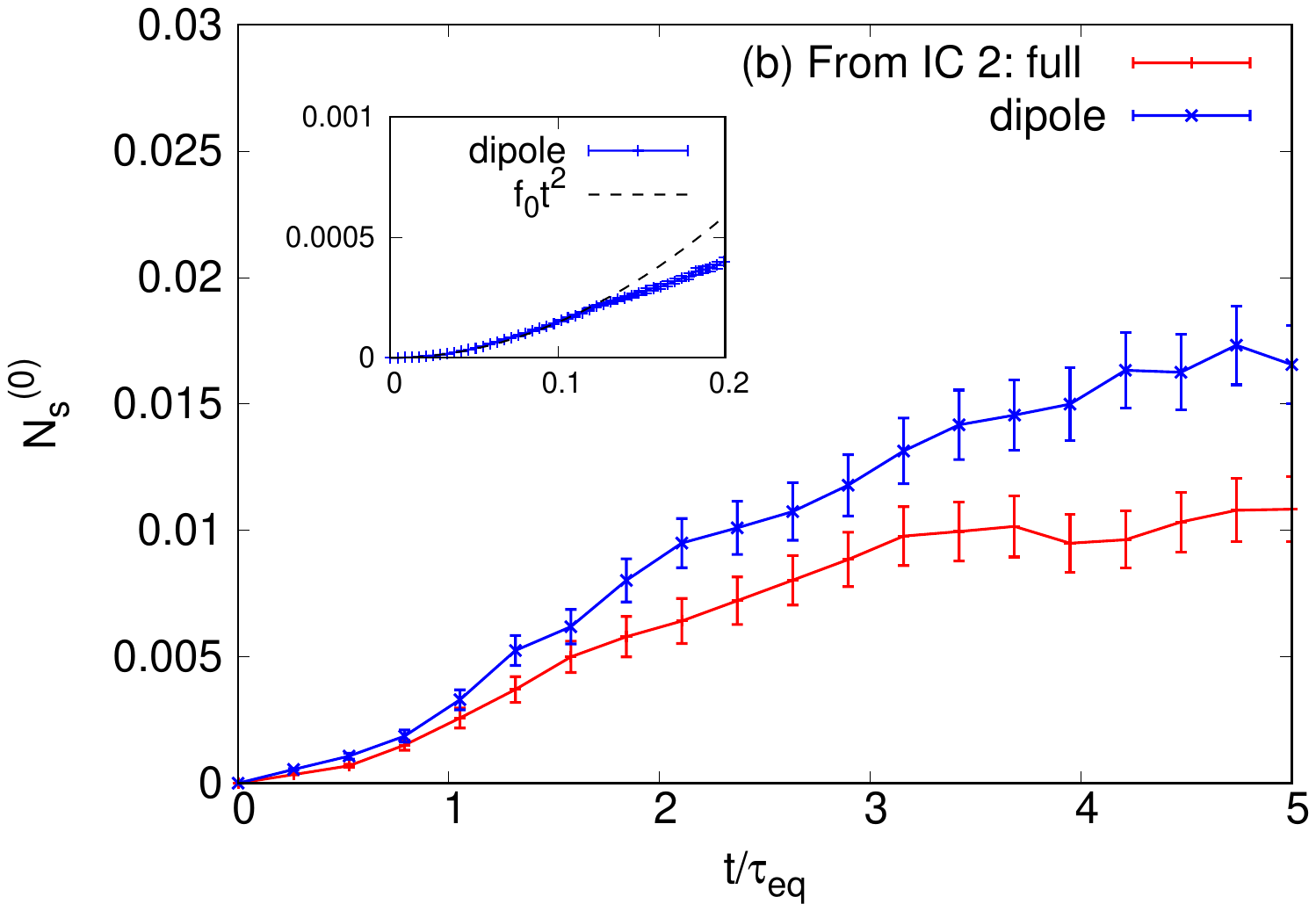}
\end{center}
\caption{Time evolution of the singlet ground state is compared with simulations with dipole-approximated Lindblad equations (a) from IC~1 and (b) from IC~2.
Short time analytical estimates are also compared.
}
\label{fig:simplification}
\end{figure}

\subsection{Applicability of dipole approximation}
In the relativistic heavy-ion collisions, the typical lifetime of the QGP fireball is $\tau_{\rm hic}\sim 10{\rm fm} \sim 250/M$ for bottom quark mass $M\approx 4.8{\rm GeV}$.
This time scale is comparable to the equilibration time for bottom quarks $\tau_{\rm eq}\simeq 236/M$ at $T=0.1M$ and is shorter than $\tau_{\rm eq}$ at lower temperatures because $\tau_{\rm eq}\propto 1/T^2$.
In such time scales, as we show below, the dipole approximation turns out to be applicable for the ground state occupation number $N_s^{(0)}(t)$.
In Fig.~\ref{fig:simplification}, we compare the simulations of the Lindblad model \eqref{eq:Lindblad} and its dipole limit \eqref{eq:Lindblad_dipole_HC} for $\gamma_2$ only ($U(r)$ is not approximated by the small $r$ expansion).
Semi-quantitative agreement is found until $t=5\tau_{\rm eq}$ or longer from IC~1 (Fig.~\ref{fig:simplification}(a)) and for $t\lesssim 1.5\tau_{\rm eq}$ from IC~2 (Fig.~\ref{fig:simplification}(b)).

From IC~1, the initial ground state is excited to the octet by the Lindblad operator $C_+(k)$ or $C_+$.
There are two effects that allow us to neglect the probability of returning back to the singlet ground state ---
(i) The octet pair repels each other and the wave function gets extended so that the spatial overlap to the ground state becomes smaller, and
(ii) The probability of flipping octet to singlet is suppressed by $\sim 1/N_c^2=1/9$ at large $N_c$ compared to that into an octet \cite{Escobedo:2020tuc, Akamatsu:2021vsh}.
Therefore, the time evolution of $N_s^{(0)}(t)$ is mostly governed by the excitation process by $C_+(k)$, whose coupling to a small ground state is well approximated by $C_+$.
To confirm this interpretation, we also compare the numerical results and the decay formula $N_s^{(0)}(t)=e^{-\Gamma_0 t}$ with
\begin{subequations}
\label{eq:Gamma_0}
\begin{align}
\Gamma_0&= \frac{-4\gamma_2}{3} \langle \psi_s^{(0)}|V_+^{\dagger}V_+ |\psi_s^{(0)}\rangle, \\
V_+ &= x - \frac{3\alpha}{8T}\frac{x}{|x|_{\rm reg}} + \frac{ip}{2MT},
\end{align}
\end{subequations}
which only takes into account the excitation process by $C_+$.
The numerical value for $\Gamma_0\simeq 1.71 \times 10^{-3}M$ is obtained for $T=0.1M$.
The quantitative agreement strongly supports our interpretation.
Furthermore, it also suggests that simulating the Schr\"odinger equation with the non-Hermitian effective Hamiltonian $H_{\rm eff}$
\begin{align}
\label{eq:effective_Hamiltonian}
H_{\rm eff} = \frac{p^2}{M} + \langle s|\Bigl[\Delta H -\frac{i}{2}\sum_{j=x,y,z}C^{\dagger}_{+j}C_{+j}\Bigr]|s\rangle,
\end{align}
which is derived by \eqref{eq:Lindblad_dipole_HC}, would be sufficient for the calculation of survival probabilities of singlet bound states.
This is an extension of the simulation of the Schr\"odinger equation with complex potential \cite{Islam:2020gdv, Katz:2022fpb, Dong:2022mbo} by including the subleading terms from the heavy quark recoil.

From IC~2, the octet wave packet feeds down to the singlet while evolving by the repulsive force and diffusion.
In short times, the evolution of the octet wave packet can be accurately described even with the dipole approximation.
This is the reason why the dipole approximation seems to work in this case even though the octet wave function expands rapidly.
To get an analytic estimate for a very short time, we calculate by the following two-step approach.
Since $C_-$ flips the parity of the wave function, the octet wave function must contain the parity odd component.
First, the octet wave packet is disturbed by $C_d$, which flips the parity, and the parity odd $(P=-1)$ octet component increases linearly in time by $\rho_o(\Delta t)_{P=-1}\simeq \Delta t C_d\rho_o(0)C_d^{\dagger}$.
Then, it is this component that yields the ground state probability after its feed down to the singlet by $C_-$.
The occupation number $N_s^{(0)}(t)$ evolves with
\begin{subequations}
\begin{align}
&\frac{dN_s^{(0)}}{dt} \simeq t \frac{5\gamma_2^2}{72}|\langle \psi_s^{(0)}|V_-V_d|\psi_o^{\rm wp}\rangle|^2 \equiv 2f_0t, \\
&V_d =x + \frac{ip}{2MT}, \quad
V_- =x + \frac{3\alpha }{8T}\frac{x}{|x|_{\rm reg}} + \frac{ip}{2MT},
\end{align}
\end{subequations}
and the short-time behavior is expected to be quadratic in time $N_s^{(0)}(t) = f_0t^2$ with $f_0=2.66\times 10^{-7} M^2$.
The quantitative agreement is found only for a short time $t\lesssim 0.12\tau_{\rm eq}$, showing that the linear approximation $\rho_o(\Delta t)_{P=-1}\propto\Delta t$ is not valid anymore at longer times.
Since the Schr\"odinger equation with \eqref{eq:effective_Hamiltonian} cannot describe the octet-to-singlet transitions and the analytical estimate of the transition rate as above has limited applicability, the dipole approximation is probably the only alternative method to solve the Lindblad equation for this kind of initial condition.

\section{Conclusion}\label{sec:Conclusion}
In this paper, we simulated the Lindblad equations for quarkonia in the QGP.
From our one-dimensional simulation of Eq.~\eqref{eq:Lindblad}, we show the equilibration process of a quarkonium with the non-Abelian color charges for the first time.
The equilibrium distribution matches the Boltzmann distribution thanks to the quantum dissipation from the recoil terms in the Lindblad equation.
We also check the applicability of the dipole approximation \eqref{eq:Lindblad_dipole_HC} by comparing two simulations with and without the approximation and found the quantitative agreement at short time scales as is realized in the relativistic heavy-ion collisions.
The simulation with the dipole approximation at a longer time scale is not reported in this paper because unphysical localization of the wave functions is found at the boundaries and thus the equilibration is not confirmed.
This is due to the mismatch between the periodic boundary condition and the linear approximation of the Lindblad operators $C_n\propto x$.
In a short enough time and a large enough volume, our analysis supports the nontrivial utility of the quarkonium Lindblad equation in the dipole limit, whose full-scale simulation has just started for the relativistic heavy-ion collision experiments \cite{Brambilla:2020qwo, Brambilla:2021wkt, Brambilla:2022ynh}.

\acknowledgements
The work of Y.A. is supported by JSPS KAKENHI Grant Number JP18K13538.
Y.A. thanks RIKEN iTHEMS Non-Equilibrium Working group (NEW) for fruitful discussions.
M.A. is supported in part by JSPS KAKENHI Grant Number JP18K03646 and JP21H00124.

\appendix
\section{Quantum State Diffusion (QSD) method}\label{app:QSD}
A Lindblad equation
\begin{align}
\frac{d}{dt} \rho(t) = -i[H,\rho] + \sum_{k}\Bigr(L_k \rho L_k^{\dagger}
-\frac{1}{2}\left\{L_k^{\dagger}L_k,\rho(t) \right\}\Bigr)
\label{eq:lindeq}
\end{align}
can be simulated by generating a wave function ensemble by solving a stochastic evolution equation.
The density matrix is reconstructed by taking a mixed state average of the wave functions
\begin{align}
\rho(t) = \lim_{N_{\rm ev}\to\infty}\frac{1}{N_{\rm ev}}\sum_{i=1}^{N_{\rm ev}}
\ket{\psi_i(t)}\bra{\psi_i(t)}.
\end{align}
Here $N_{\rm ev}$ is a number of trials by the stochastic evolution and $|\psi_i(t)\rangle \ (i=1, 2,  \cdots, N_{\rm ev})$ is a wave function of the $i$-th event at time $t$.
In the limit $N_{\rm ev}\to \infty$, the average defined above equals the solution of the Lindblad equation.
In general, this kind of method is called ``stochastic unravelling" and another well-known method is called the quantum jump \cite{Plenio:1997ep} and is used in \cite{Brambilla:2020qwo, Brambilla:2021wkt, Brambilla:2022ynh}.

In the QSD method \cite{gisin1992quantum}, wave functions evolve according to the following nonlinear stochastic Schr\"{o}dinger equation,
\begin{align}
\ket{d\psi}&\equiv\ket{\psi(t+dt)}-\ket{\psi(t)}
\notag\\
&=-iH\ket{\psi(t)}dt \nonumber \\
&\quad +\sum_{k}\Bigr(\langle L^{\dagger}_k\rangle_{\psi} L_k-\frac{1}{2}L^{\dagger}_k L_k-\frac{1}{2}\langle L^{\dagger}_k \rangle_{\psi}\langle L_k \rangle_{\psi}
\Bigr)\ket{\psi(t)}dt
\notag\\
&\quad +\sum_k \Bigr(L_k- \langle L_k \rangle_{\psi}\Bigr)\ket{\psi(t)}d\xi_k\,,
\end{align}
and $d\xi_k$ is {\it complex} white noise satisfying
\begin{subequations}
\begin{align}
&\langle d\xi_k\rangle = \langle\Re(d\xi_k)\Im(d\xi_l)\rangle=0, \\
&\langle \Re(d\xi_k)\Re(d\xi_l)\rangle =\langle\Im(d\xi_k)\Im(d\xi_l)\rangle=\delta_{kl} dt/2.
\end{align}
\end{subequations}
The nonlinearity arises from the terms $\langle L_k^{(\dagger)}\rangle_{\psi} = \bra{\psi}L_k^{(\dagger)}\ket{\psi}$.
Since some terms only play the role of norm conservation, one can drop them to get a stochastic evolution equation for unnormalized wave functions
\begin{align}
\ket{d\phi(t)}
&=-iH\ket{\phi(t)}dt \nonumber \\
&\quad+\sum_{k}\bigr(\langle L^{\dagger}_k \rangle_{\phi} L_k - \frac{1}{2}L^{\dagger}_k L_k\bigr)\ket{\phi(t)}dt \nonumber \\
& \quad+\sum_k L_k \ket{\phi(t)}d\xi_k,
\end{align}
which we employ in our numerical simulation.
Note that in this case $\langle L_k^{\dagger}\rangle_{\phi} = \bra{\phi}L_k^{\dagger}\ket{\phi}/\langle\phi|\phi\rangle$.
The density matrix is obtained by
\begin{align}
\rho(t) &= \lim_{N_{\rm ev}\to\infty}\frac{1}{N_{\rm ev}}\sum_{i=1}^{N_{\rm ev}}
\frac{\ket{\phi_i(t)}\bra{\phi_i(t)}}{\langle\phi_i(t)|\phi_i(t)\rangle}.
\end{align}
The actual numerical implementation of quarkonium Lindblad equation \eqref{eq:Lindblad} is described in detail in \cite{MiuraThesisOsakaU}.

\bibliography{paper_v10}

\begin{thebibliography}{61}%
\makeatletter
\providecommand \@ifxundefined [1]{%
 \@ifx{#1\undefined}
}%
\providecommand \@ifnum [1]{%
 \ifnum #1\expandafter \@firstoftwo
 \else \expandafter \@secondoftwo
 \fi
}%
\providecommand \@ifx [1]{%
 \ifx #1\expandafter \@firstoftwo
 \else \expandafter \@secondoftwo
 \fi
}%
\providecommand \natexlab [1]{#1}%
\providecommand \enquote  [1]{``#1''}%
\providecommand \bibnamefont  [1]{#1}%
\providecommand \bibfnamefont [1]{#1}%
\providecommand \citenamefont [1]{#1}%
\providecommand \href@noop [0]{\@secondoftwo}%
\providecommand \href [0]{\begingroup \@sanitize@url \@href}%
\providecommand \@href[1]{\@@startlink{#1}\@@href}%
\providecommand \@@href[1]{\endgroup#1\@@endlink}%
\providecommand \@sanitize@url [0]{\catcode `\\12\catcode `\$12\catcode
  `\&12\catcode `\#12\catcode `\^12\catcode `\_12\catcode `\%12\relax}%
\providecommand \@@startlink[1]{}%
\providecommand \@@endlink[0]{}%
\providecommand \url  [0]{\begingroup\@sanitize@url \@url }%
\providecommand \@url [1]{\endgroup\@href {#1}{\urlprefix }}%
\providecommand \urlprefix  [0]{URL }%
\providecommand \Eprint [0]{\href }%
\providecommand \doibase [0]{https://doi.org/}%
\providecommand \selectlanguage [0]{\@gobble}%
\providecommand \bibinfo  [0]{\@secondoftwo}%
\providecommand \bibfield  [0]{\@secondoftwo}%
\providecommand \translation [1]{[#1]}%
\providecommand \BibitemOpen [0]{}%
\providecommand \bibitemStop [0]{}%
\providecommand \bibitemNoStop [0]{.\EOS\space}%
\providecommand \EOS [0]{\spacefactor3000\relax}%
\providecommand \BibitemShut  [1]{\csname bibitem#1\endcsname}%
\let\auto@bib@innerbib\@empty
\bibitem [{\citenamefont {Yagi}\ \emph {et~al.}(2005)\citenamefont {Yagi},
  \citenamefont {Hatsuda},\ and\ \citenamefont {Miake}}]{Yagi:985487}%
  \BibitemOpen
  \bibfield  {author} {\bibinfo {author} {\bibfnamefont {K.}~\bibnamefont
  {Yagi}}, \bibinfo {author} {\bibfnamefont {T.}~\bibnamefont {Hatsuda}},\ and\
  \bibinfo {author} {\bibfnamefont {Y.}~\bibnamefont {Miake}},\ }\href@noop {}
  {\emph {\bibinfo {title} {Quark-Gluon Plasma: From Big Bang to Little
  Bang}}}\ (\bibinfo {address} {Cambridge},\ \bibinfo {year}
  {2005})\BibitemShut {NoStop}%
\bibitem [{\citenamefont {Bernhard}\ \emph {et~al.}(2019)\citenamefont
  {Bernhard}, \citenamefont {Moreland},\ and\ \citenamefont
  {Bass}}]{Bernhard:2019bmu}%
  \BibitemOpen
  \bibfield  {author} {\bibinfo {author} {\bibfnamefont {J.~E.}\ \bibnamefont
  {Bernhard}}, \bibinfo {author} {\bibfnamefont {J.~S.}\ \bibnamefont
  {Moreland}},\ and\ \bibinfo {author} {\bibfnamefont {S.~A.}\ \bibnamefont
  {Bass}},\ }\bibfield  {title} {\bibinfo {title} {{Bayesian estimation of the
  specific shear and bulk viscosity of quark\textendash{}gluon plasma}},\
  }\href {https://doi.org/10.1038/s41567-019-0611-8} {\bibfield  {journal}
  {\bibinfo  {journal} {Nature Phys.}\ }\textbf {\bibinfo {volume} {15}},\
  \bibinfo {pages} {1113} (\bibinfo {year} {2019})}\BibitemShut {NoStop}%
\bibitem [{\citenamefont {Chatrchyan}\ \emph {et~al.}(2011)\citenamefont
  {Chatrchyan} \emph {et~al.}}]{Chatrchyan:2011pe}%
  \BibitemOpen
  \bibfield  {author} {\bibinfo {author} {\bibfnamefont {S.}~\bibnamefont
  {Chatrchyan}} \emph {et~al.} (\bibinfo {collaboration} {CMS}),\ }\bibfield
  {title} {\bibinfo {title} {{Indications of suppression of excited $\Upsilon$
  states in PbPb collisions at $\sqrt{S_{NN}}$ = 2.76 TeV}},\ }\href
  {https://doi.org/10.1103/PhysRevLett.107.052302} {\bibfield  {journal}
  {\bibinfo  {journal} {Phys. Rev. Lett.}\ }\textbf {\bibinfo {volume} {107}},\
  \bibinfo {pages} {052302} (\bibinfo {year} {2011})},\ \Eprint
  {https://arxiv.org/abs/1105.4894} {arXiv:1105.4894 [nucl-ex]} \BibitemShut
  {NoStop}%
\bibitem [{\citenamefont {Chatrchyan}\ \emph {et~al.}(2012)\citenamefont
  {Chatrchyan} \emph {et~al.}}]{Chatrchyan:2012lxa}%
  \BibitemOpen
  \bibfield  {author} {\bibinfo {author} {\bibfnamefont {S.}~\bibnamefont
  {Chatrchyan}} \emph {et~al.} (\bibinfo {collaboration} {CMS}),\ }\bibfield
  {title} {\bibinfo {title} {{Observation of sequential Upsilon suppression in
  PbPb collisions}},\ }\href {https://doi.org/10.1103/PhysRevLett.109.222301}
  {\bibfield  {journal} {\bibinfo  {journal} {Phys. Rev. Lett.}\ }\textbf
  {\bibinfo {volume} {109}},\ \bibinfo {pages} {222301} (\bibinfo {year}
  {2012})},\ \Eprint {https://arxiv.org/abs/1208.2826} {arXiv:1208.2826
  [nucl-ex]} \BibitemShut {NoStop}%
\bibitem [{\citenamefont {Khachatryan}\ \emph {et~al.}(2016)\citenamefont
  {Khachatryan} \emph {et~al.}}]{Khachatryan:2016xxp}%
  \BibitemOpen
  \bibfield  {author} {\bibinfo {author} {\bibfnamefont {V.}~\bibnamefont
  {Khachatryan}} \emph {et~al.} (\bibinfo {collaboration} {CMS}),\ }\bibfield
  {title} {\bibinfo {title} {{Suppression of $\Upsilon(1S)$, $\Upsilon(2S)$ and
  $\Upsilon(3S)$ production in PbPb collisions at $\sqrt{s_{NN}}$ = 2.76
  TeV}},\ }\href@noop {} {\bibfield  {journal} {\bibinfo  {journal} {Submitted
  to: Phys. Lett. B}\ } (\bibinfo {year} {2016})},\ \Eprint
  {https://arxiv.org/abs/1611.01510} {arXiv:1611.01510 [nucl-ex]} \BibitemShut
  {NoStop}%
\bibitem [{\citenamefont {Breuer}\ and\ \citenamefont
  {Petruccione}(2002)}]{BRE02}%
  \BibitemOpen
  \bibfield  {author} {\bibinfo {author} {\bibfnamefont {H.~P.}\ \bibnamefont
  {Breuer}}\ and\ \bibinfo {author} {\bibfnamefont {F.}~\bibnamefont
  {Petruccione}},\ }\href@noop {} {\emph {\bibinfo {title} {The theory of open
  quantum systems}}}\ (\bibinfo  {publisher} {Oxford University Press},\
  \bibinfo {year} {2002})\BibitemShut {NoStop}%
\bibitem [{\citenamefont {Rivas}\ and\ \citenamefont
  {Huelga}(2012)}]{rivas2012open}%
  \BibitemOpen
  \bibfield  {author} {\bibinfo {author} {\bibfnamefont {A.}~\bibnamefont
  {Rivas}}\ and\ \bibinfo {author} {\bibfnamefont {S.~F.}\ \bibnamefont
  {Huelga}},\ }\href@noop {} {\emph {\bibinfo {title} {Open quantum
  systems}}},\ Vol.~\bibinfo {volume} {10}\ (\bibinfo  {publisher} {Springer},\
  \bibinfo {year} {2012})\BibitemShut {NoStop}%
\bibitem [{\citenamefont {Laine}\ \emph {et~al.}(2007)\citenamefont {Laine},
  \citenamefont {Philipsen}, \citenamefont {Romatschke},\ and\ \citenamefont
  {Tassler}}]{Laine:2006ns}%
  \BibitemOpen
  \bibfield  {author} {\bibinfo {author} {\bibfnamefont {M.}~\bibnamefont
  {Laine}}, \bibinfo {author} {\bibfnamefont {O.}~\bibnamefont {Philipsen}},
  \bibinfo {author} {\bibfnamefont {P.}~\bibnamefont {Romatschke}},\ and\
  \bibinfo {author} {\bibfnamefont {M.}~\bibnamefont {Tassler}},\ }\bibfield
  {title} {\bibinfo {title} {{Real-time static potential in hot QCD}},\ }\href
  {https://doi.org/10.1088/1126-6708/2007/03/054} {\bibfield  {journal}
  {\bibinfo  {journal} {JHEP}\ }\textbf {\bibinfo {volume} {03}},\ \bibinfo
  {pages} {054}},\ \Eprint {https://arxiv.org/abs/hep-ph/0611300}
  {arXiv:hep-ph/0611300 [hep-ph]} \BibitemShut {NoStop}%
\bibitem [{\citenamefont {Beraudo}\ \emph {et~al.}(2008)\citenamefont
  {Beraudo}, \citenamefont {Blaizot},\ and\ \citenamefont
  {Ratti}}]{Beraudo:2007ky}%
  \BibitemOpen
  \bibfield  {author} {\bibinfo {author} {\bibfnamefont {A.}~\bibnamefont
  {Beraudo}}, \bibinfo {author} {\bibfnamefont {J.~P.}\ \bibnamefont
  {Blaizot}},\ and\ \bibinfo {author} {\bibfnamefont {C.}~\bibnamefont
  {Ratti}},\ }\bibfield  {title} {\bibinfo {title} {{Real and imaginary-time Q
  anti-Q correlators in a thermal medium}},\ }\href
  {https://doi.org/10.1016/j.nuclphysa.2008.03.001} {\bibfield  {journal}
  {\bibinfo  {journal} {Nucl. Phys.}\ }\textbf {\bibinfo {volume} {A806}},\
  \bibinfo {pages} {312} (\bibinfo {year} {2008})},\ \Eprint
  {https://arxiv.org/abs/0712.4394} {arXiv:0712.4394 [nucl-th]} \BibitemShut
  {NoStop}%
\bibitem [{\citenamefont {Brambilla}\ \emph {et~al.}(2008)\citenamefont
  {Brambilla}, \citenamefont {Ghiglieri}, \citenamefont {Vairo},\ and\
  \citenamefont {Petreczky}}]{Brambilla:2008cx}%
  \BibitemOpen
  \bibfield  {author} {\bibinfo {author} {\bibfnamefont {N.}~\bibnamefont
  {Brambilla}}, \bibinfo {author} {\bibfnamefont {J.}~\bibnamefont
  {Ghiglieri}}, \bibinfo {author} {\bibfnamefont {A.}~\bibnamefont {Vairo}},\
  and\ \bibinfo {author} {\bibfnamefont {P.}~\bibnamefont {Petreczky}},\
  }\bibfield  {title} {\bibinfo {title} {{Static quark-antiquark pairs at
  finite temperature}},\ }\href {https://doi.org/10.1103/PhysRevD.78.014017}
  {\bibfield  {journal} {\bibinfo  {journal} {Phys. Rev.}\ }\textbf {\bibinfo
  {volume} {D78}},\ \bibinfo {pages} {014017} (\bibinfo {year} {2008})},\
  \Eprint {https://arxiv.org/abs/0804.0993} {arXiv:0804.0993 [hep-ph]}
  \BibitemShut {NoStop}%
\bibitem [{\citenamefont {Rothkopf}\ \emph {et~al.}(2012)\citenamefont
  {Rothkopf}, \citenamefont {Hatsuda},\ and\ \citenamefont
  {Sasaki}}]{Rothkopf:2011db}%
  \BibitemOpen
  \bibfield  {author} {\bibinfo {author} {\bibfnamefont {A.}~\bibnamefont
  {Rothkopf}}, \bibinfo {author} {\bibfnamefont {T.}~\bibnamefont {Hatsuda}},\
  and\ \bibinfo {author} {\bibfnamefont {S.}~\bibnamefont {Sasaki}},\
  }\bibfield  {title} {\bibinfo {title} {{Complex Heavy-Quark Potential at
  Finite Temperature from Lattice QCD}},\ }\href
  {https://doi.org/10.1103/PhysRevLett.108.162001} {\bibfield  {journal}
  {\bibinfo  {journal} {Phys. Rev. Lett.}\ }\textbf {\bibinfo {volume} {108}},\
  \bibinfo {pages} {162001} (\bibinfo {year} {2012})},\ \Eprint
  {https://arxiv.org/abs/1108.1579} {arXiv:1108.1579 [hep-lat]} \BibitemShut
  {NoStop}%
\bibitem [{\citenamefont {Young}\ and\ \citenamefont
  {Dusling}(2013)}]{Young:2010jq}%
  \BibitemOpen
  \bibfield  {author} {\bibinfo {author} {\bibfnamefont {C.}~\bibnamefont
  {Young}}\ and\ \bibinfo {author} {\bibfnamefont {K.}~\bibnamefont
  {Dusling}},\ }\bibfield  {title} {\bibinfo {title} {{Quarkonium above
  deconfinement as an open quantum system}},\ }\href
  {https://doi.org/10.1103/PhysRevC.87.065206} {\bibfield  {journal} {\bibinfo
  {journal} {Phys. Rev.}\ }\textbf {\bibinfo {volume} {C87}},\ \bibinfo {pages}
  {065206} (\bibinfo {year} {2013})},\ \Eprint
  {https://arxiv.org/abs/1001.0935} {arXiv:1001.0935 [nucl-th]} \BibitemShut
  {NoStop}%
\bibitem [{\citenamefont {Borghini}\ and\ \citenamefont
  {Gombeaud}(2011)}]{Borghini:2011yq}%
  \BibitemOpen
  \bibfield  {author} {\bibinfo {author} {\bibfnamefont {N.}~\bibnamefont
  {Borghini}}\ and\ \bibinfo {author} {\bibfnamefont {C.}~\bibnamefont
  {Gombeaud}},\ }\bibfield  {title} {\bibinfo {title} {{Dynamical Evolution of
  Heavy Quarkonia in a Deconfined Medium}},\ }\href@noop {} {\  (\bibinfo
  {year} {2011})},\ \Eprint {https://arxiv.org/abs/1103.2945} {arXiv:1103.2945
  [hep-ph]} \BibitemShut {NoStop}%
\bibitem [{\citenamefont {Borghini}\ and\ \citenamefont
  {Gombeaud}(2012)}]{Borghini:2011ms}%
  \BibitemOpen
  \bibfield  {author} {\bibinfo {author} {\bibfnamefont {N.}~\bibnamefont
  {Borghini}}\ and\ \bibinfo {author} {\bibfnamefont {C.}~\bibnamefont
  {Gombeaud}},\ }\bibfield  {title} {\bibinfo {title} {{Heavy quarkonia in a
  medium as a quantum dissipative system: Master equation approach}},\ }\href
  {https://doi.org/10.1140/epjc/s10052-012-2000-7} {\bibfield  {journal}
  {\bibinfo  {journal} {Eur. Phys. J.}\ }\textbf {\bibinfo {volume} {C72}},\
  \bibinfo {pages} {2000} (\bibinfo {year} {2012})},\ \Eprint
  {https://arxiv.org/abs/1109.4271} {arXiv:1109.4271 [nucl-th]} \BibitemShut
  {NoStop}%
\bibitem [{\citenamefont {Akamatsu}\ and\ \citenamefont
  {Rothkopf}(2012)}]{Akamatsu:2011se}%
  \BibitemOpen
  \bibfield  {author} {\bibinfo {author} {\bibfnamefont {Y.}~\bibnamefont
  {Akamatsu}}\ and\ \bibinfo {author} {\bibfnamefont {A.}~\bibnamefont
  {Rothkopf}},\ }\bibfield  {title} {\bibinfo {title} {{Stochastic potential
  and quantum decoherence of heavy quarkonium in the quark-gluon plasma}},\
  }\href {https://doi.org/10.1103/PhysRevD.85.105011} {\bibfield  {journal}
  {\bibinfo  {journal} {Phys. Rev. D}\ }\textbf {\bibinfo {volume} {85}},\
  \bibinfo {pages} {105011} (\bibinfo {year} {2012})},\ \Eprint
  {https://arxiv.org/abs/1110.1203} {arXiv:1110.1203 [hep-ph]} \BibitemShut
  {NoStop}%
\bibitem [{\citenamefont {Blaizot}\ \emph {et~al.}(2016)\citenamefont
  {Blaizot}, \citenamefont {De~Boni}, \citenamefont {Faccioli},\ and\
  \citenamefont {Garberoglio}}]{Blaizot:2015hya}%
  \BibitemOpen
  \bibfield  {author} {\bibinfo {author} {\bibfnamefont {J.-P.}\ \bibnamefont
  {Blaizot}}, \bibinfo {author} {\bibfnamefont {D.}~\bibnamefont {De~Boni}},
  \bibinfo {author} {\bibfnamefont {P.}~\bibnamefont {Faccioli}},\ and\
  \bibinfo {author} {\bibfnamefont {G.}~\bibnamefont {Garberoglio}},\
  }\bibfield  {title} {\bibinfo {title} {{Heavy quark bound states in a
  quark-gluon plasma: Dissociation and recombination}},\ }\href
  {https://doi.org/10.1016/j.nuclphysa.2015.10.011} {\bibfield  {journal}
  {\bibinfo  {journal} {Nucl. Phys.}\ }\textbf {\bibinfo {volume} {A946}},\
  \bibinfo {pages} {49} (\bibinfo {year} {2016})},\ \Eprint
  {https://arxiv.org/abs/1503.03857} {arXiv:1503.03857 [nucl-th]} \BibitemShut
  {NoStop}%
\bibitem [{\citenamefont {Blaizot}\ and\ \citenamefont
  {Escobedo}(2018{\natexlab{a}})}]{Blaizot:2017ypk}%
  \BibitemOpen
  \bibfield  {author} {\bibinfo {author} {\bibfnamefont {J.-P.}\ \bibnamefont
  {Blaizot}}\ and\ \bibinfo {author} {\bibfnamefont {M.~A.}\ \bibnamefont
  {Escobedo}},\ }\bibfield  {title} {\bibinfo {title} {{Quantum and classical
  dynamics of heavy quarks in a quark-gluon plasma}},\ }\href
  {https://doi.org/10.1007/JHEP06(2018)034} {\bibfield  {journal} {\bibinfo
  {journal} {JHEP}\ }\textbf {\bibinfo {volume} {06}},\ \bibinfo {pages}
  {034}},\ \Eprint {https://arxiv.org/abs/1711.10812} {arXiv:1711.10812
  [hep-ph]} \BibitemShut {NoStop}%
\bibitem [{\citenamefont {Blaizot}\ and\ \citenamefont
  {Escobedo}(2018{\natexlab{b}})}]{Blaizot:2018oev}%
  \BibitemOpen
  \bibfield  {author} {\bibinfo {author} {\bibfnamefont {J.-P.}\ \bibnamefont
  {Blaizot}}\ and\ \bibinfo {author} {\bibfnamefont {M.~A.}\ \bibnamefont
  {Escobedo}},\ }\bibfield  {title} {\bibinfo {title} {{Approach to equilibrium
  of a quarkonium in a quark-gluon plasma}},\ }\href
  {https://doi.org/10.1103/PhysRevD.98.074007} {\bibfield  {journal} {\bibinfo
  {journal} {Phys. Rev. D}\ }\textbf {\bibinfo {volume} {98}},\ \bibinfo
  {pages} {074007} (\bibinfo {year} {2018}{\natexlab{b}})},\ \Eprint
  {https://arxiv.org/abs/1803.07996} {arXiv:1803.07996 [hep-ph]} \BibitemShut
  {NoStop}%
\bibitem [{\citenamefont {Lindblad}(1976)}]{Lindblad:1975ef}%
  \BibitemOpen
  \bibfield  {author} {\bibinfo {author} {\bibfnamefont {G.}~\bibnamefont
  {Lindblad}},\ }\bibfield  {title} {\bibinfo {title} {{On the Generators of
  Quantum Dynamical Semigroups}},\ }\href {https://doi.org/10.1007/BF01608499}
  {\bibfield  {journal} {\bibinfo  {journal} {Commun. Math. Phys.}\ }\textbf
  {\bibinfo {volume} {48}},\ \bibinfo {pages} {119} (\bibinfo {year}
  {1976})}\BibitemShut {NoStop}%
\bibitem [{\citenamefont {Gorini}\ \emph {et~al.}(1976)\citenamefont {Gorini},
  \citenamefont {Kossakowski},\ and\ \citenamefont
  {Sudarshan}}]{Gorini:1975nb}%
  \BibitemOpen
  \bibfield  {author} {\bibinfo {author} {\bibfnamefont {V.}~\bibnamefont
  {Gorini}}, \bibinfo {author} {\bibfnamefont {A.}~\bibnamefont
  {Kossakowski}},\ and\ \bibinfo {author} {\bibfnamefont {E.~C.~G.}\
  \bibnamefont {Sudarshan}},\ }\bibfield  {title} {\bibinfo {title}
  {{Completely Positive Dynamical Semigroups of N Level Systems}},\ }\href
  {https://doi.org/10.1063/1.522979} {\bibfield  {journal} {\bibinfo  {journal}
  {J. Math. Phys.}\ }\textbf {\bibinfo {volume} {17}},\ \bibinfo {pages} {821}
  (\bibinfo {year} {1976})}\BibitemShut {NoStop}%
\bibitem [{\citenamefont {Akamatsu}(2015)}]{Akamatsu:2014qsa}%
  \BibitemOpen
  \bibfield  {author} {\bibinfo {author} {\bibfnamefont {Y.}~\bibnamefont
  {Akamatsu}},\ }\bibfield  {title} {\bibinfo {title} {{Heavy quark master
  equations in the Lindblad form at high temperatures}},\ }\href
  {https://doi.org/10.1103/PhysRevD.91.056002} {\bibfield  {journal} {\bibinfo
  {journal} {Phys. Rev. D}\ }\textbf {\bibinfo {volume} {91}},\ \bibinfo
  {pages} {056002} (\bibinfo {year} {2015})},\ \Eprint
  {https://arxiv.org/abs/1403.5783} {arXiv:1403.5783 [hep-ph]} \BibitemShut
  {NoStop}%
\bibitem [{\citenamefont {Brambilla}\ \emph {et~al.}(2017)\citenamefont
  {Brambilla}, \citenamefont {Escobedo}, \citenamefont {Soto},\ and\
  \citenamefont {Vairo}}]{Brambilla:2016wgg}%
  \BibitemOpen
  \bibfield  {author} {\bibinfo {author} {\bibfnamefont {N.}~\bibnamefont
  {Brambilla}}, \bibinfo {author} {\bibfnamefont {M.~A.}\ \bibnamefont
  {Escobedo}}, \bibinfo {author} {\bibfnamefont {J.}~\bibnamefont {Soto}},\
  and\ \bibinfo {author} {\bibfnamefont {A.}~\bibnamefont {Vairo}},\ }\bibfield
   {title} {\bibinfo {title} {{Quarkonium suppression in heavy-ion collisions:
  an open quantum system approach}},\ }\href
  {https://doi.org/10.1103/PhysRevD.96.034021} {\bibfield  {journal} {\bibinfo
  {journal} {Phys. Rev.}\ }\textbf {\bibinfo {volume} {D96}},\ \bibinfo {pages}
  {034021} (\bibinfo {year} {2017})},\ \Eprint
  {https://arxiv.org/abs/1612.07248} {arXiv:1612.07248 [hep-ph]} \BibitemShut
  {NoStop}%
\bibitem [{\citenamefont {Brambilla}\ \emph {et~al.}(2018)\citenamefont
  {Brambilla}, \citenamefont {Escobedo}, \citenamefont {Soto},\ and\
  \citenamefont {Vairo}}]{Brambilla:2017zei}%
  \BibitemOpen
  \bibfield  {author} {\bibinfo {author} {\bibfnamefont {N.}~\bibnamefont
  {Brambilla}}, \bibinfo {author} {\bibfnamefont {M.~A.}\ \bibnamefont
  {Escobedo}}, \bibinfo {author} {\bibfnamefont {J.}~\bibnamefont {Soto}},\
  and\ \bibinfo {author} {\bibfnamefont {A.}~\bibnamefont {Vairo}},\ }\bibfield
   {title} {\bibinfo {title} {{Heavy quarkonium suppression in a fireball}},\
  }\href {https://doi.org/10.1103/PhysRevD.97.074009} {\bibfield  {journal}
  {\bibinfo  {journal} {Phys. Rev. D}\ }\textbf {\bibinfo {volume} {97}},\
  \bibinfo {pages} {074009} (\bibinfo {year} {2018})},\ \Eprint
  {https://arxiv.org/abs/1711.04515} {arXiv:1711.04515 [hep-ph]} \BibitemShut
  {NoStop}%
\bibitem [{\citenamefont {Yao}\ and\ \citenamefont
  {Mehen}(2019)}]{Yao:2018nmy}%
  \BibitemOpen
  \bibfield  {author} {\bibinfo {author} {\bibfnamefont {X.}~\bibnamefont
  {Yao}}\ and\ \bibinfo {author} {\bibfnamefont {T.}~\bibnamefont {Mehen}},\
  }\bibfield  {title} {\bibinfo {title} {{Quarkonium in-medium transport
  equation derived from first principles}},\ }\href
  {https://doi.org/10.1103/PhysRevD.99.096028} {\bibfield  {journal} {\bibinfo
  {journal} {Phys. Rev. D}\ }\textbf {\bibinfo {volume} {99}},\ \bibinfo
  {pages} {096028} (\bibinfo {year} {2019})},\ \Eprint
  {https://arxiv.org/abs/1811.07027} {arXiv:1811.07027 [hep-ph]} \BibitemShut
  {NoStop}%
\bibitem [{\citenamefont {Rothkopf}(2014)}]{Rothkopf:2013kya}%
  \BibitemOpen
  \bibfield  {author} {\bibinfo {author} {\bibfnamefont {A.}~\bibnamefont
  {Rothkopf}},\ }\bibfield  {title} {\bibinfo {title} {{A first look at
  Bottomonium melting via a stochastic potential}},\ }\href
  {https://doi.org/10.1007/JHEP04(2014)085} {\bibfield  {journal} {\bibinfo
  {journal} {JHEP}\ }\textbf {\bibinfo {volume} {04}},\ \bibinfo {pages}
  {085}},\ \Eprint {https://arxiv.org/abs/1312.3246} {arXiv:1312.3246 [hep-ph]}
  \BibitemShut {NoStop}%
\bibitem [{\citenamefont {Kajimoto}\ \emph {et~al.}(2018)\citenamefont
  {Kajimoto}, \citenamefont {Akamatsu}, \citenamefont {Asakawa},\ and\
  \citenamefont {Rothkopf}}]{Kajimoto:2017rel}%
  \BibitemOpen
  \bibfield  {author} {\bibinfo {author} {\bibfnamefont {S.}~\bibnamefont
  {Kajimoto}}, \bibinfo {author} {\bibfnamefont {Y.}~\bibnamefont {Akamatsu}},
  \bibinfo {author} {\bibfnamefont {M.}~\bibnamefont {Asakawa}},\ and\ \bibinfo
  {author} {\bibfnamefont {A.}~\bibnamefont {Rothkopf}},\ }\bibfield  {title}
  {\bibinfo {title} {{Dynamical dissociation of quarkonia by wave function
  decoherence}},\ }\href {https://doi.org/10.1103/PhysRevD.97.014003}
  {\bibfield  {journal} {\bibinfo  {journal} {Phys. Rev. D}\ }\textbf {\bibinfo
  {volume} {97}},\ \bibinfo {pages} {014003} (\bibinfo {year} {2018})},\
  \Eprint {https://arxiv.org/abs/1705.03365} {arXiv:1705.03365 [nucl-th]}
  \BibitemShut {NoStop}%
\bibitem [{\citenamefont {Akamatsu}\ \emph {et~al.}(2018)\citenamefont
  {Akamatsu}, \citenamefont {Asakawa}, \citenamefont {Kajimoto},\ and\
  \citenamefont {Rothkopf}}]{Akamatsu:2018xim}%
  \BibitemOpen
  \bibfield  {author} {\bibinfo {author} {\bibfnamefont {Y.}~\bibnamefont
  {Akamatsu}}, \bibinfo {author} {\bibfnamefont {M.}~\bibnamefont {Asakawa}},
  \bibinfo {author} {\bibfnamefont {S.}~\bibnamefont {Kajimoto}},\ and\
  \bibinfo {author} {\bibfnamefont {A.}~\bibnamefont {Rothkopf}},\ }\bibfield
  {title} {\bibinfo {title} {{Quantum dissipation of a heavy quark from a
  nonlinear stochastic Schr\"odinger equation}},\ }\href
  {https://doi.org/10.1007/JHEP07(2018)029} {\bibfield  {journal} {\bibinfo
  {journal} {JHEP}\ }\textbf {\bibinfo {volume} {07}},\ \bibinfo {pages}
  {029}},\ \Eprint {https://arxiv.org/abs/1805.00167} {arXiv:1805.00167
  [nucl-th]} \BibitemShut {NoStop}%
\bibitem [{\citenamefont {Miura}\ \emph {et~al.}(2020)\citenamefont {Miura},
  \citenamefont {Akamatsu}, \citenamefont {Asakawa},\ and\ \citenamefont
  {Rothkopf}}]{Miura:2019ssi}%
  \BibitemOpen
  \bibfield  {author} {\bibinfo {author} {\bibfnamefont {T.}~\bibnamefont
  {Miura}}, \bibinfo {author} {\bibfnamefont {Y.}~\bibnamefont {Akamatsu}},
  \bibinfo {author} {\bibfnamefont {M.}~\bibnamefont {Asakawa}},\ and\ \bibinfo
  {author} {\bibfnamefont {A.}~\bibnamefont {Rothkopf}},\ }\bibfield  {title}
  {\bibinfo {title} {{Quantum Brownian motion of a heavy quark pair in the
  quark-gluon plasma}},\ }\href {https://doi.org/10.1103/PhysRevD.101.034011}
  {\bibfield  {journal} {\bibinfo  {journal} {Phys. Rev. D}\ }\textbf {\bibinfo
  {volume} {101}},\ \bibinfo {pages} {034011} (\bibinfo {year} {2020})},\
  \Eprint {https://arxiv.org/abs/1908.06293} {arXiv:1908.06293 [nucl-th]}
  \BibitemShut {NoStop}%
\bibitem [{\citenamefont {\r{A}lund}\ \emph {et~al.}(2021)\citenamefont
  {\r{A}lund}, \citenamefont {Akamatsu}, \citenamefont {Laur\'en},
  \citenamefont {Miura}, \citenamefont {Nordstr\"om},\ and\ \citenamefont
  {Rothkopf}}]{Alund:2020ctu}%
  \BibitemOpen
  \bibfield  {author} {\bibinfo {author} {\bibfnamefont {O.}~\bibnamefont
  {\r{A}lund}}, \bibinfo {author} {\bibfnamefont {Y.}~\bibnamefont {Akamatsu}},
  \bibinfo {author} {\bibfnamefont {F.}~\bibnamefont {Laur\'en}}, \bibinfo
  {author} {\bibfnamefont {T.}~\bibnamefont {Miura}}, \bibinfo {author}
  {\bibfnamefont {J.}~\bibnamefont {Nordstr\"om}},\ and\ \bibinfo {author}
  {\bibfnamefont {A.}~\bibnamefont {Rothkopf}},\ }\bibfield  {title} {\bibinfo
  {title} {{Trace preserving quantum dynamics using a novel
  reparametrization-neutral summation-by-parts difference operator}},\ }\href
  {https://doi.org/10.1016/j.jcp.2020.109917} {\bibfield  {journal} {\bibinfo
  {journal} {J. Comput. Phys.}\ }\textbf {\bibinfo {volume} {425}},\ \bibinfo
  {pages} {109917} (\bibinfo {year} {2021})},\ \Eprint
  {https://arxiv.org/abs/2004.04406} {arXiv:2004.04406 [physics.comp-ph]}
  \BibitemShut {NoStop}%
\bibitem [{\citenamefont {Akamatsu}\ \emph {et~al.}(2022)\citenamefont
  {Akamatsu}, \citenamefont {Asakawa},\ and\ \citenamefont
  {Kajimoto}}]{Akamatsu:2021vsh}%
  \BibitemOpen
  \bibfield  {author} {\bibinfo {author} {\bibfnamefont {Y.}~\bibnamefont
  {Akamatsu}}, \bibinfo {author} {\bibfnamefont {M.}~\bibnamefont {Asakawa}},\
  and\ \bibinfo {author} {\bibfnamefont {S.}~\bibnamefont {Kajimoto}},\
  }\bibfield  {title} {\bibinfo {title} {{Dynamics of in-medium quarkonia in
  SU(3) and SU(2) gauge theories}},\ }\href
  {https://doi.org/10.1103/PhysRevD.105.054036} {\bibfield  {journal} {\bibinfo
   {journal} {Phys. Rev. D}\ }\textbf {\bibinfo {volume} {105}},\ \bibinfo
  {pages} {054036} (\bibinfo {year} {2022})},\ \Eprint
  {https://arxiv.org/abs/2108.06921} {arXiv:2108.06921 [nucl-th]} \BibitemShut
  {NoStop}%
\bibitem [{\citenamefont {Brambilla}\ \emph
  {et~al.}(2021{\natexlab{a}})\citenamefont {Brambilla}, \citenamefont
  {Escobedo}, \citenamefont {Strickland}, \citenamefont {Vairo}, \citenamefont
  {Vander~Griend},\ and\ \citenamefont {Weber}}]{Brambilla:2020qwo}%
  \BibitemOpen
  \bibfield  {author} {\bibinfo {author} {\bibfnamefont {N.}~\bibnamefont
  {Brambilla}}, \bibinfo {author} {\bibfnamefont {M.~A.}\ \bibnamefont
  {Escobedo}}, \bibinfo {author} {\bibfnamefont {M.}~\bibnamefont
  {Strickland}}, \bibinfo {author} {\bibfnamefont {A.}~\bibnamefont {Vairo}},
  \bibinfo {author} {\bibfnamefont {P.}~\bibnamefont {Vander~Griend}},\ and\
  \bibinfo {author} {\bibfnamefont {J.~H.}\ \bibnamefont {Weber}},\ }\bibfield
  {title} {\bibinfo {title} {{Bottomonium suppression in an open quantum system
  using the quantum trajectories method}},\ }\href
  {https://doi.org/10.1007/JHEP05(2021)136} {\bibfield  {journal} {\bibinfo
  {journal} {JHEP}\ }\textbf {\bibinfo {volume} {05}},\ \bibinfo {pages}
  {136}},\ \Eprint {https://arxiv.org/abs/2012.01240} {arXiv:2012.01240
  [hep-ph]} \BibitemShut {NoStop}%
\bibitem [{\citenamefont {Brambilla}\ \emph
  {et~al.}(2021{\natexlab{b}})\citenamefont {Brambilla}, \citenamefont
  {Escobedo}, \citenamefont {Strickland}, \citenamefont {Vairo}, \citenamefont
  {Griend},\ and\ \citenamefont {Weber}}]{Brambilla:2021wkt}%
  \BibitemOpen
  \bibfield  {author} {\bibinfo {author} {\bibfnamefont {N.}~\bibnamefont
  {Brambilla}}, \bibinfo {author} {\bibfnamefont {M.~A.}\ \bibnamefont
  {Escobedo}}, \bibinfo {author} {\bibfnamefont {M.}~\bibnamefont
  {Strickland}}, \bibinfo {author} {\bibfnamefont {A.}~\bibnamefont {Vairo}},
  \bibinfo {author} {\bibfnamefont {P.~V.}\ \bibnamefont {Griend}},\ and\
  \bibinfo {author} {\bibfnamefont {J.~H.}\ \bibnamefont {Weber}},\ }\bibfield
  {title} {\bibinfo {title} {{Bottomonium production in heavy-ion collisions
  using quantum trajectories: Differential observables and momentum
  anisotropy}},\ }\href@noop {} {\  (\bibinfo {year} {2021}{\natexlab{b}})},\
  \Eprint {https://arxiv.org/abs/2107.06222} {arXiv:2107.06222 [hep-ph]}
  \BibitemShut {NoStop}%
\bibitem [{\citenamefont {Brambilla}\ \emph {et~al.}(2022)\citenamefont
  {Brambilla}, \citenamefont {Escobedo}, \citenamefont {Islam}, \citenamefont
  {Strickland}, \citenamefont {Tiwari}, \citenamefont {Vairo},\ and\
  \citenamefont {Vander~Griend}}]{Brambilla:2022ynh}%
  \BibitemOpen
  \bibfield  {author} {\bibinfo {author} {\bibfnamefont {N.}~\bibnamefont
  {Brambilla}}, \bibinfo {author} {\bibfnamefont {M.~A.}\ \bibnamefont
  {Escobedo}}, \bibinfo {author} {\bibfnamefont {A.}~\bibnamefont {Islam}},
  \bibinfo {author} {\bibfnamefont {M.}~\bibnamefont {Strickland}}, \bibinfo
  {author} {\bibfnamefont {A.}~\bibnamefont {Tiwari}}, \bibinfo {author}
  {\bibfnamefont {A.}~\bibnamefont {Vairo}},\ and\ \bibinfo {author}
  {\bibfnamefont {P.}~\bibnamefont {Vander~Griend}},\ }\bibfield  {title}
  {\bibinfo {title} {{Heavy quarkonium dynamics at next-to-leading order in the
  binding energy over temperature}},\ }\href@noop {} {\  (\bibinfo {year}
  {2022})},\ \Eprint {https://arxiv.org/abs/2205.10289} {arXiv:2205.10289
  [hep-ph]} \BibitemShut {NoStop}%
\bibitem [{\citenamefont {Sharma}\ and\ \citenamefont
  {Tiwari}(2020)}]{Sharma:2019xum}%
  \BibitemOpen
  \bibfield  {author} {\bibinfo {author} {\bibfnamefont {R.}~\bibnamefont
  {Sharma}}\ and\ \bibinfo {author} {\bibfnamefont {A.}~\bibnamefont
  {Tiwari}},\ }\bibfield  {title} {\bibinfo {title} {{Quantum evolution of
  quarkonia with correlated and uncorrelated noise}},\ }\href
  {https://doi.org/10.1103/PhysRevD.101.074004} {\bibfield  {journal} {\bibinfo
   {journal} {Phys. Rev. D}\ }\textbf {\bibinfo {volume} {101}},\ \bibinfo
  {pages} {074004} (\bibinfo {year} {2020})},\ \Eprint
  {https://arxiv.org/abs/1912.07036} {arXiv:1912.07036 [hep-ph]} \BibitemShut
  {NoStop}%
\bibitem [{\citenamefont {Yao}\ \emph {et~al.}(2021)\citenamefont {Yao},
  \citenamefont {Ke}, \citenamefont {Xu}, \citenamefont {Bass},\ and\
  \citenamefont {M\"uller}}]{Yao:2020xzw}%
  \BibitemOpen
  \bibfield  {author} {\bibinfo {author} {\bibfnamefont {X.}~\bibnamefont
  {Yao}}, \bibinfo {author} {\bibfnamefont {W.}~\bibnamefont {Ke}}, \bibinfo
  {author} {\bibfnamefont {Y.}~\bibnamefont {Xu}}, \bibinfo {author}
  {\bibfnamefont {S.~A.}\ \bibnamefont {Bass}},\ and\ \bibinfo {author}
  {\bibfnamefont {B.}~\bibnamefont {M\"uller}},\ }\bibfield  {title} {\bibinfo
  {title} {{Coupled Boltzmann Transport Equations of Heavy Quarks and Quarkonia
  in Quark-Gluon Plasma}},\ }\href {https://doi.org/10.1007/JHEP01(2021)046}
  {\bibfield  {journal} {\bibinfo  {journal} {JHEP}\ }\textbf {\bibinfo
  {volume} {01}},\ \bibinfo {pages} {046}},\ \Eprint
  {https://arxiv.org/abs/2004.06746} {arXiv:2004.06746 [hep-ph]} \BibitemShut
  {NoStop}%
\bibitem [{\citenamefont {Akamatsu}(2022)}]{Akamatsu:2020ypb}%
  \BibitemOpen
  \bibfield  {author} {\bibinfo {author} {\bibfnamefont {Y.}~\bibnamefont
  {Akamatsu}},\ }\bibfield  {title} {\bibinfo {title} {{Quarkonium in
  quark\textendash{}gluon plasma: Open quantum system approaches
  re-examined}},\ }\href {https://doi.org/10.1016/j.ppnp.2021.103932}
  {\bibfield  {journal} {\bibinfo  {journal} {Prog. Part. Nucl. Phys.}\
  }\textbf {\bibinfo {volume} {123}},\ \bibinfo {pages} {103932} (\bibinfo
  {year} {2022})},\ \Eprint {https://arxiv.org/abs/2009.10559}
  {arXiv:2009.10559 [nucl-th]} \BibitemShut {NoStop}%
\bibitem [{\citenamefont {Yao}(2021)}]{Yao:2021lus}%
  \BibitemOpen
  \bibfield  {author} {\bibinfo {author} {\bibfnamefont {X.}~\bibnamefont
  {Yao}},\ }\bibfield  {title} {\bibinfo {title} {{Open quantum systems for
  quarkonia}},\ }\href {https://doi.org/10.1142/S0217751X21300106} {\bibfield
  {journal} {\bibinfo  {journal} {Int. J. Mod. Phys. A}\ }\textbf {\bibinfo
  {volume} {36}},\ \bibinfo {pages} {2130010} (\bibinfo {year} {2021})},\
  \Eprint {https://arxiv.org/abs/2102.01736} {arXiv:2102.01736 [hep-ph]}
  \BibitemShut {NoStop}%
\bibitem [{\citenamefont {Sharma}(2021)}]{Sharma:2021vvu}%
  \BibitemOpen
  \bibfield  {author} {\bibinfo {author} {\bibfnamefont {R.}~\bibnamefont
  {Sharma}},\ }\bibfield  {title} {\bibinfo {title} {{Quarkonium propagation in
  the quark\textendash{}gluon plasma}},\ }\href
  {https://doi.org/10.1140/epjs/s11734-021-00025-z} {\bibfield  {journal}
  {\bibinfo  {journal} {Eur. Phys. J. ST}\ }\textbf {\bibinfo {volume} {230}},\
  \bibinfo {pages} {697} (\bibinfo {year} {2021})},\ \Eprint
  {https://arxiv.org/abs/2101.04268} {arXiv:2101.04268 [hep-ph]} \BibitemShut
  {NoStop}%
\bibitem [{\citenamefont {Zhao}\ and\ \citenamefont
  {Rapp}(2011)}]{Zhao:2011cv}%
  \BibitemOpen
  \bibfield  {author} {\bibinfo {author} {\bibfnamefont {X.}~\bibnamefont
  {Zhao}}\ and\ \bibinfo {author} {\bibfnamefont {R.}~\bibnamefont {Rapp}},\
  }\bibfield  {title} {\bibinfo {title} {{Medium Modifications and Production
  of Charmonia at LHC}},\ }\href
  {https://doi.org/10.1016/j.nuclphysa.2011.05.001} {\bibfield  {journal}
  {\bibinfo  {journal} {Nucl. Phys. A}\ }\textbf {\bibinfo {volume} {859}},\
  \bibinfo {pages} {114} (\bibinfo {year} {2011})},\ \Eprint
  {https://arxiv.org/abs/1102.2194} {arXiv:1102.2194 [hep-ph]} \BibitemShut
  {NoStop}%
\bibitem [{\citenamefont {Song}\ \emph {et~al.}(2011)\citenamefont {Song},
  \citenamefont {Han},\ and\ \citenamefont {Ko}}]{Song:2011xi}%
  \BibitemOpen
  \bibfield  {author} {\bibinfo {author} {\bibfnamefont {T.}~\bibnamefont
  {Song}}, \bibinfo {author} {\bibfnamefont {K.~C.}\ \bibnamefont {Han}},\ and\
  \bibinfo {author} {\bibfnamefont {C.~M.}\ \bibnamefont {Ko}},\ }\bibfield
  {title} {\bibinfo {title} {{Charmonium production in relativistic heavy-ion
  collisions}},\ }\href {https://doi.org/10.1103/PhysRevC.84.034907} {\bibfield
   {journal} {\bibinfo  {journal} {Phys. Rev. C}\ }\textbf {\bibinfo {volume}
  {84}},\ \bibinfo {pages} {034907} (\bibinfo {year} {2011})},\ \Eprint
  {https://arxiv.org/abs/1103.6197} {arXiv:1103.6197 [nucl-th]} \BibitemShut
  {NoStop}%
\bibitem [{\citenamefont {Zhou}\ \emph {et~al.}(2014)\citenamefont {Zhou},
  \citenamefont {Xu}, \citenamefont {Xu},\ and\ \citenamefont
  {Zhuang}}]{Zhou:2014kka}%
  \BibitemOpen
  \bibfield  {author} {\bibinfo {author} {\bibfnamefont {K.}~\bibnamefont
  {Zhou}}, \bibinfo {author} {\bibfnamefont {N.}~\bibnamefont {Xu}}, \bibinfo
  {author} {\bibfnamefont {Z.}~\bibnamefont {Xu}},\ and\ \bibinfo {author}
  {\bibfnamefont {P.}~\bibnamefont {Zhuang}},\ }\bibfield  {title} {\bibinfo
  {title} {{Medium effects on charmonium production at ultrarelativistic
  energies available at the CERN Large Hadron Collider}},\ }\href
  {https://doi.org/10.1103/PhysRevC.89.054911} {\bibfield  {journal} {\bibinfo
  {journal} {Phys. Rev. C}\ }\textbf {\bibinfo {volume} {89}},\ \bibinfo
  {pages} {054911} (\bibinfo {year} {2014})},\ \Eprint
  {https://arxiv.org/abs/1401.5845} {arXiv:1401.5845 [nucl-th]} \BibitemShut
  {NoStop}%
\bibitem [{\citenamefont {Du}\ and\ \citenamefont {Rapp}(2022)}]{Du:2022uvj}%
  \BibitemOpen
  \bibfield  {author} {\bibinfo {author} {\bibfnamefont {X.}~\bibnamefont
  {Du}}\ and\ \bibinfo {author} {\bibfnamefont {R.}~\bibnamefont {Rapp}},\
  }\bibfield  {title} {\bibinfo {title} {{Non-equilibrium charmonium
  regeneration in strongly coupled quark-gluon plasma}},\ }\href@noop {} {\
  (\bibinfo {year} {2022})},\ \Eprint {https://arxiv.org/abs/2207.00065}
  {arXiv:2207.00065 [nucl-th]} \BibitemShut {NoStop}%
\bibitem [{\citenamefont {Abelev}\ \emph {et~al.}(2014)\citenamefont {Abelev}
  \emph {et~al.}}]{Abelev:2013ila}%
  \BibitemOpen
  \bibfield  {author} {\bibinfo {author} {\bibfnamefont {B.~B.}\ \bibnamefont
  {Abelev}} \emph {et~al.} (\bibinfo {collaboration} {ALICE}),\ }\bibfield
  {title} {\bibinfo {title} {{Centrality, rapidity and transverse momentum
  dependence of $J/\psi$ suppression in Pb-Pb collisions at $\sqrt{s_{\rm
  NN}}$=2.76 TeV}},\ }\href {https://doi.org/10.1016/j.physletb.2014.05.064}
  {\bibfield  {journal} {\bibinfo  {journal} {Phys. Lett.}\ }\textbf {\bibinfo
  {volume} {B734}},\ \bibinfo {pages} {314} (\bibinfo {year} {2014})},\ \Eprint
  {https://arxiv.org/abs/1311.0214} {arXiv:1311.0214 [nucl-ex]} \BibitemShut
  {NoStop}%
\bibitem [{\citenamefont {Adam}\ \emph {et~al.}(2017)\citenamefont {Adam} \emph
  {et~al.}}]{Adam:2016rdg}%
  \BibitemOpen
  \bibfield  {author} {\bibinfo {author} {\bibfnamefont {J.}~\bibnamefont
  {Adam}} \emph {et~al.} (\bibinfo {collaboration} {ALICE}),\ }\bibfield
  {title} {\bibinfo {title} {{J/$\psi$ suppression at forward rapidity in Pb-Pb
  collisions at $\mathbf{\sqrt{s_{{\rm NN}}} = 5.02}$ TeV}},\ }\href
  {https://doi.org/10.1016/j.physletb.2016.12.064} {\bibfield  {journal}
  {\bibinfo  {journal} {Phys. Lett.}\ }\textbf {\bibinfo {volume} {B766}},\
  \bibinfo {pages} {212} (\bibinfo {year} {2017})},\ \Eprint
  {https://arxiv.org/abs/1606.08197} {arXiv:1606.08197 [nucl-ex]} \BibitemShut
  {NoStop}%
\bibitem [{\citenamefont {Akamatsu}\ and\ \citenamefont
  {Miura}(2022)}]{Akamatsu:2021dot}%
  \BibitemOpen
  \bibfield  {author} {\bibinfo {author} {\bibfnamefont {Y.}~\bibnamefont
  {Akamatsu}}\ and\ \bibinfo {author} {\bibfnamefont {T.}~\bibnamefont
  {Miura}},\ }\bibfield  {title} {\bibinfo {title} {{Nonequilibrium evolution
  of quarkonium in medium}},\ }\href
  {https://doi.org/10.1051/epjconf/202225801006} {\bibfield  {journal}
  {\bibinfo  {journal} {EPJ Web Conf.}\ }\textbf {\bibinfo {volume} {258}},\
  \bibinfo {pages} {01006} (\bibinfo {year} {2022})},\ \Eprint
  {https://arxiv.org/abs/2111.15402} {arXiv:2111.15402 [hep-ph]} \BibitemShut
  {NoStop}%
\bibitem [{\citenamefont {Spohn}(1977)}]{spohn1977algebraic}%
  \BibitemOpen
  \bibfield  {author} {\bibinfo {author} {\bibfnamefont {H.}~\bibnamefont
  {Spohn}},\ }\bibfield  {title} {\bibinfo {title} {An algebraic condition for
  the approach to equilibrium of an open n-level system},\ }\href
  {https://doi.org/10.1007/BF00420668} {\bibfield  {journal} {\bibinfo
  {journal} {Letters in Mathematical Physics}\ }\textbf {\bibinfo {volume}
  {2}},\ \bibinfo {pages} {33} (\bibinfo {year} {1977})}\BibitemShut {NoStop}%
\bibitem [{\citenamefont {Schirmer}\ and\ \citenamefont
  {Wang}(2010)}]{schirmer2010stabilizing}%
  \BibitemOpen
  \bibfield  {author} {\bibinfo {author} {\bibfnamefont {S.}~\bibnamefont
  {Schirmer}}\ and\ \bibinfo {author} {\bibfnamefont {X.}~\bibnamefont
  {Wang}},\ }\bibfield  {title} {\bibinfo {title} {Stabilizing open quantum
  systems by markovian reservoir engineering},\ }\href
  {https://doi.org/10.1103/PhysRevA.81.062306} {\bibfield  {journal} {\bibinfo
  {journal} {Physical Review A}\ }\textbf {\bibinfo {volume} {81}},\ \bibinfo
  {pages} {062306} (\bibinfo {year} {2010})}\BibitemShut {NoStop}%
\bibitem [{\citenamefont {Caldeira}\ and\ \citenamefont
  {Leggett}(1983)}]{Caldeira:1982iu}%
  \BibitemOpen
  \bibfield  {author} {\bibinfo {author} {\bibfnamefont {A.~O.}\ \bibnamefont
  {Caldeira}}\ and\ \bibinfo {author} {\bibfnamefont {A.~J.}\ \bibnamefont
  {Leggett}},\ }\bibfield  {title} {\bibinfo {title} {{Path integral approach
  to quantum Brownian motion}},\ }\href
  {https://doi.org/10.1016/0378-4371(83)90013-4} {\bibfield  {journal}
  {\bibinfo  {journal} {Physica A}\ }\textbf {\bibinfo {volume} {121}},\
  \bibinfo {pages} {587} (\bibinfo {year} {1983})}\BibitemShut {NoStop}%
\bibitem [{\citenamefont {Moore}\ and\ \citenamefont
  {Teaney}(2005)}]{Moore:2004tg}%
  \BibitemOpen
  \bibfield  {author} {\bibinfo {author} {\bibfnamefont {G.~D.}\ \bibnamefont
  {Moore}}\ and\ \bibinfo {author} {\bibfnamefont {D.}~\bibnamefont {Teaney}},\
  }\bibfield  {title} {\bibinfo {title} {{How much do heavy quarks thermalize
  in a heavy ion collision?}},\ }\href
  {https://doi.org/10.1103/PhysRevC.71.064904} {\bibfield  {journal} {\bibinfo
  {journal} {Phys. Rev. C}\ }\textbf {\bibinfo {volume} {71}},\ \bibinfo
  {pages} {064904} (\bibinfo {year} {2005})},\ \Eprint
  {https://arxiv.org/abs/hep-ph/0412346} {arXiv:hep-ph/0412346} \BibitemShut
  {NoStop}%
\bibitem [{\citenamefont {Bali}(2001)}]{Bali:2000gf}%
  \BibitemOpen
  \bibfield  {author} {\bibinfo {author} {\bibfnamefont {G.~S.}\ \bibnamefont
  {Bali}},\ }\bibfield  {title} {\bibinfo {title} {{QCD forces and heavy quark
  bound states}},\ }\href {https://doi.org/10.1016/S0370-1573(00)00079-X}
  {\bibfield  {journal} {\bibinfo  {journal} {Phys. Rept.}\ }\textbf {\bibinfo
  {volume} {343}},\ \bibinfo {pages} {1} (\bibinfo {year} {2001})},\ \Eprint
  {https://arxiv.org/abs/hep-ph/0001312} {arXiv:hep-ph/0001312} \BibitemShut
  {NoStop}%
\bibitem [{\citenamefont {Alford}\ and\ \citenamefont
  {Strickland}(2013)}]{Alford:2013jva}%
  \BibitemOpen
  \bibfield  {author} {\bibinfo {author} {\bibfnamefont {J.}~\bibnamefont
  {Alford}}\ and\ \bibinfo {author} {\bibfnamefont {M.}~\bibnamefont
  {Strickland}},\ }\bibfield  {title} {\bibinfo {title} {{Charmonia and
  Bottomonia in a Magnetic Field}},\ }\href
  {https://doi.org/10.1103/PhysRevD.88.105017} {\bibfield  {journal} {\bibinfo
  {journal} {Phys. Rev. D}\ }\textbf {\bibinfo {volume} {88}},\ \bibinfo
  {pages} {105017} (\bibinfo {year} {2013})},\ \Eprint
  {https://arxiv.org/abs/1309.3003} {arXiv:1309.3003 [hep-ph]} \BibitemShut
  {NoStop}%
\bibitem [{\citenamefont {Gisin}\ and\ \citenamefont
  {Percival}(1992)}]{gisin1992quantum}%
  \BibitemOpen
  \bibfield  {author} {\bibinfo {author} {\bibfnamefont {N.}~\bibnamefont
  {Gisin}}\ and\ \bibinfo {author} {\bibfnamefont {I.~C.}\ \bibnamefont
  {Percival}},\ }\bibfield  {title} {\bibinfo {title} {The quantum-state
  diffusion model applied to open systems},\ }\href
  {https://doi.org/10.1088/0305-4470/25/21/023} {\bibfield  {journal} {\bibinfo
   {journal} {Journal of Physics A: Mathematical and General}\ }\textbf
  {\bibinfo {volume} {25}},\ \bibinfo {pages} {5677} (\bibinfo {year}
  {1992})}\BibitemShut {NoStop}%
\bibitem [{\citenamefont {Miura}(2021)}]{MiuraThesisOsakaU}%
  \BibitemOpen
  \bibfield  {author} {\bibinfo {author} {\bibfnamefont {T.}~\bibnamefont
  {Miura}},\ }\emph {\bibinfo {title} {{Quantum dissipation of quarkonium in
  the quark-gluon plasma via Lindblad equation}}},\ \href@noop {} {Ph.D.
  thesis},\ \bibinfo  {school} {{Osaka University}} (\bibinfo {year}
  {{2021}})\BibitemShut {NoStop}%
\bibitem [{\citenamefont
  {Casalderrey-Solana}(2013)}]{Casalderrey-Solana:2012yfo}%
  \BibitemOpen
  \bibfield  {author} {\bibinfo {author} {\bibfnamefont {J.}~\bibnamefont
  {Casalderrey-Solana}},\ }\bibfield  {title} {\bibinfo {title} {{Dynamical
  Quarkonia Suppression in a QGP-Brick}},\ }\href
  {https://doi.org/10.1007/JHEP03(2013)091} {\bibfield  {journal} {\bibinfo
  {journal} {JHEP}\ }\textbf {\bibinfo {volume} {03}},\ \bibinfo {pages}
  {091}},\ \Eprint {https://arxiv.org/abs/1208.2602} {arXiv:1208.2602 [hep-ph]}
  \BibitemShut {NoStop}%
\bibitem [{\citenamefont {Cho}\ and\ \citenamefont
  {Leibovich}(1996{\natexlab{a}})}]{Cho:1995vh}%
  \BibitemOpen
  \bibfield  {author} {\bibinfo {author} {\bibfnamefont {P.~L.}\ \bibnamefont
  {Cho}}\ and\ \bibinfo {author} {\bibfnamefont {A.~K.}\ \bibnamefont
  {Leibovich}},\ }\bibfield  {title} {\bibinfo {title} {{Color octet quarkonia
  production}},\ }\href {https://doi.org/10.1103/PhysRevD.53.150} {\bibfield
  {journal} {\bibinfo  {journal} {Phys. Rev. D}\ }\textbf {\bibinfo {volume}
  {53}},\ \bibinfo {pages} {150} (\bibinfo {year} {1996}{\natexlab{a}})},\
  \Eprint {https://arxiv.org/abs/hep-ph/9505329} {arXiv:hep-ph/9505329}
  \BibitemShut {NoStop}%
\bibitem [{\citenamefont {Cho}\ and\ \citenamefont
  {Leibovich}(1996{\natexlab{b}})}]{Cho:1995ce}%
  \BibitemOpen
  \bibfield  {author} {\bibinfo {author} {\bibfnamefont {P.~L.}\ \bibnamefont
  {Cho}}\ and\ \bibinfo {author} {\bibfnamefont {A.~K.}\ \bibnamefont
  {Leibovich}},\ }\bibfield  {title} {\bibinfo {title} {{Color octet quarkonia
  production. 2.}},\ }\href {https://doi.org/10.1103/PhysRevD.53.6203}
  {\bibfield  {journal} {\bibinfo  {journal} {Phys. Rev. D}\ }\textbf {\bibinfo
  {volume} {53}},\ \bibinfo {pages} {6203} (\bibinfo {year}
  {1996}{\natexlab{b}})},\ \Eprint {https://arxiv.org/abs/hep-ph/9511315}
  {arXiv:hep-ph/9511315} \BibitemShut {NoStop}%
\bibitem [{\citenamefont {Escobedo}(2021)}]{Escobedo:2020tuc}%
  \BibitemOpen
  \bibfield  {author} {\bibinfo {author} {\bibfnamefont {M.~A.}\ \bibnamefont
  {Escobedo}},\ }\bibfield  {title} {\bibinfo {title} {{Medium evolution of a
  static quark-antiquark pair in the large $N_c$ limit}},\ }\href
  {https://doi.org/10.1103/PhysRevD.103.034010} {\bibfield  {journal} {\bibinfo
   {journal} {Phys. Rev. D}\ }\textbf {\bibinfo {volume} {103}},\ \bibinfo
  {pages} {034010} (\bibinfo {year} {2021})},\ \Eprint
  {https://arxiv.org/abs/2010.10424} {arXiv:2010.10424 [hep-ph]} \BibitemShut
  {NoStop}%
\bibitem [{\citenamefont {Islam}\ and\ \citenamefont
  {Strickland}(2020)}]{Islam:2020gdv}%
  \BibitemOpen
  \bibfield  {author} {\bibinfo {author} {\bibfnamefont {A.}~\bibnamefont
  {Islam}}\ and\ \bibinfo {author} {\bibfnamefont {M.}~\bibnamefont
  {Strickland}},\ }\bibfield  {title} {\bibinfo {title} {{Bottomonium
  suppression and elliptic flow from real-time quantum evolution}},\ }\href
  {https://doi.org/10.1016/j.physletb.2020.135949} {\bibfield  {journal}
  {\bibinfo  {journal} {Phys. Lett. B}\ }\textbf {\bibinfo {volume} {811}},\
  \bibinfo {pages} {135949} (\bibinfo {year} {2020})},\ \Eprint
  {https://arxiv.org/abs/2007.10211} {arXiv:2007.10211 [hep-ph]} \BibitemShut
  {NoStop}%
\bibitem [{\citenamefont {Katz}\ \emph {et~al.}(2022)\citenamefont {Katz},
  \citenamefont {Delorme},\ and\ \citenamefont {Gossiaux}}]{Katz:2022fpb}%
  \BibitemOpen
  \bibfield  {author} {\bibinfo {author} {\bibfnamefont {R.}~\bibnamefont
  {Katz}}, \bibinfo {author} {\bibfnamefont {S.}~\bibnamefont {Delorme}},\ and\
  \bibinfo {author} {\bibfnamefont {P.-B.}\ \bibnamefont {Gossiaux}},\
  }\bibfield  {title} {\bibinfo {title} {{One-dimensional complex potentials
  for quarkonia in a quark-gluon plasma}},\ }\href@noop {} {\  (\bibinfo {year}
  {2022})},\ \Eprint {https://arxiv.org/abs/2205.05154} {arXiv:2205.05154
  [hep-ph]} \BibitemShut {NoStop}%
\bibitem [{\citenamefont {Dong}\ \emph {et~al.}(2022)\citenamefont {Dong},
  \citenamefont {Guo}, \citenamefont {Islam}, \citenamefont {Rothkopf},\ and\
  \citenamefont {Strickland}}]{Dong:2022mbo}%
  \BibitemOpen
  \bibfield  {author} {\bibinfo {author} {\bibfnamefont {L.}~\bibnamefont
  {Dong}}, \bibinfo {author} {\bibfnamefont {Y.}~\bibnamefont {Guo}}, \bibinfo
  {author} {\bibfnamefont {A.}~\bibnamefont {Islam}}, \bibinfo {author}
  {\bibfnamefont {A.}~\bibnamefont {Rothkopf}},\ and\ \bibinfo {author}
  {\bibfnamefont {M.}~\bibnamefont {Strickland}},\ }\bibfield  {title}
  {\bibinfo {title} {{The complex heavy-quark potential in an anisotropic
  quark-gluon plasma -- Statics and dynamics}},\ }\href@noop {} {\  (\bibinfo
  {year} {2022})},\ \Eprint {https://arxiv.org/abs/2205.10349}
  {arXiv:2205.10349 [hep-ph]} \BibitemShut {NoStop}%
\bibitem [{\citenamefont {Plenio}\ and\ \citenamefont
  {Knight}(1998)}]{Plenio:1997ep}%
  \BibitemOpen
  \bibfield  {author} {\bibinfo {author} {\bibfnamefont {M.~B.}\ \bibnamefont
  {Plenio}}\ and\ \bibinfo {author} {\bibfnamefont {P.~L.}\ \bibnamefont
  {Knight}},\ }\bibfield  {title} {\bibinfo {title} {{The Quantum jump approach
  to dissipative dynamics in quantum optics}},\ }\href
  {https://doi.org/10.1103/RevModPhys.70.101} {\bibfield  {journal} {\bibinfo
  {journal} {Rev. Mod. Phys.}\ }\textbf {\bibinfo {volume} {70}},\ \bibinfo
  {pages} {101} (\bibinfo {year} {1998})},\ \Eprint
  {https://arxiv.org/abs/quant-ph/9702007} {arXiv:quant-ph/9702007}
  \BibitemShut {NoStop}%
\end{thebibliography}%
\end{document}